\documentclass[eqsecnum, aps, onecolumn, superscriptaddress, 12pt ]{revtex4-2}
\usepackage{graphicx}
\usepackage{dcolumn}
\usepackage[usenames,dvipsnames,svgnames,table]{xcolor}
\usepackage[colorlinks=true,
linkcolor=blue,
urlcolor=black,
citecolor=blue]{hyperref}
\usepackage{bm}
\usepackage{enumerate}
\usepackage{lipsum}
\usepackage{epstopdf}
\usepackage{float}
\usepackage[caption = false]{subfig}
\usepackage{tabularx}
\usepackage{color}
\usepackage{longtable}
\usepackage{rotating}
\usepackage[utf8]{inputenc}

{\tiny {\tiny }}
\begin{document}

\preprint{APS/123-QED}

\title{Two-photon  polarizability of Ba$^+$ ion: Control of spin-mixing processes in an ultracold $^{137}$Ba$^+$--$^{87}$Rb mixture}

\author{Arghya Das}
\email{arghyadas@iitkgp.ac.in}
\affiliation{Department of Physics, Indian Institute of Technology Kharagpur, Kharagpur-721302, India.}
\author{Anal Bhowmik}
\email{abhowmik@campus.haifa.ac.il}
\affiliation{Haifa Research Center for Theoretical Physics and Astrophysics, University of Haifa, Haifa 3498838, Israel}
\affiliation{Department of Mathematics, University of Haifa, Haifa 3498838, Israel}
\author{Narendra Nath Dutta}
\email{narendranathdutta7@gmail.com}
\affiliation{School of Sciences, SR University, Warangal,Telangana-506371, India.}
\author{Sonjoy Majumder}
\email{sonjoym@phy.iitkgp.ernet.in}
\affiliation{Department of Physics, Indian Institute of Technology Kharagpur, Kharagpur-721302, India.}

\date{\today}

\begin{abstract}

Ionic clocks exhibit as the most promising candidates for the frequency standards. Recent investigations show the profound advantages of interrogating two laser beams with different frequencies in developing the frequency standards. Here we present a scheme of a two-photon mechanism to calculate the dynamic polarizabilities for the clock states, 6$^2$S$_{\frac{1}{2}}$ and 5$^2$D$_{\frac{3}{2}, \frac{5}{2}}$, of Ba$^+$ by employing relativistic coupled-cluster method. We illustrate the Stark-shift cancellation between these clock states at the two-photon magic wavelengths. These magic wavelengths can be essential inputs to achieve better accuracy in the ionic clock experiments. We also calculate the magic wavelengths under the single-photon interaction to serve as the reference and for a comparative study. The calculated single- and two-photon magic wavelengths lie in the optical region and thus are significant for future state-of-the-art experiments. Moreover, as an application of the two-photon polarizabilities, we investigate the impact of these polarizabilities on the spin-mixing processes, $|0,0\rangle$ $\leftrightarrow$ $|+1,-1\rangle$ and $|0,0\rangle$ $\leftrightarrow$ $|-1,+1\rangle$, of an ultra-cold spin-1 mixture of the $^{137}$Ba$^+$ and $^{87}$Rb atoms. We determine the protocols of selecting these spin-mixing oscillations by changing the strength of an externally applied magnetic field and the frequencies of the interrogating laser beams.

\end{abstract}

\maketitle

\section{INTRODUCTION} \label{INT}

Theoretical advancements towards achieving high accuracy have led to reliable experimental explorations on optical clocks \cite{Arnold2020,Chanu2020}, processing of quantum information \cite{Inlek2017,Hucul2017}, measurements of fundamental constants, and other atomic properties \cite{Dutta2014, Kozlov2018, Hucul2017, Munshi2015, Dijck2015}. Singly charged barium atom or Ba$^+$ is one of the promising candidates for these experiments. It has well-understood energy levels and long-lived 5$^2D_{\frac{3}{2}, \frac{5}{2}}$ states as the first two excited states. This ion has been drawing the attention of several theorists and experimentalists over the last two decades or so. The recent theoretical and experimental ventures \cite{Arnold2020,Chanu2020,Zhang2020} prove the extant significance of this ion in the advances of science and technology. The laser, which can be used to control this ion for experimental purposes, shifts the energy levels of this ion due to the Stark effect. Consequently, these Stark-shifts give rise to uncertainty in the frequency measurements of the 6$^2S_{\frac{1}{2}}$ $-$ 5$^2D_{\frac{3}{2}, \frac{5}{2}}$ optical clock transitions of this ion. To eliminate this uncertainty, one needs to know the accurate dynamic or frequency-dependent profiles of the polarizabilities \cite{Mitroy2010, Tang2013} for the clock states 6$^2S_{\frac{1}{2}}$ and 5$^2D_{\frac{3}{2}, \frac{5}{2}}$. The intersections between the polarizability profiles of the 6$^2S_{\frac{1}{2}}$ and 5$^2D_{\frac{3}{2}, \frac{5}{2}}$ states provide the values of the magic wavelengths \cite{Mitroy2010, Tang2013}. If the wavelength of the laser light is tuned to a magic wavelength, the differential Stark-shift between the clock states vanishes. Moreover, the static scalar part of the differential polarizability between the clock states can give a measure of the black-body radiation (BBR) shift \cite{Gallagher1979, Porsev2006, Chanu2020} of the clock frequency. The BBR shift is reported as one of the most significant systematic uncertainty for the present-day International System of Units (SI) \cite{Jefferts2014}.

\begin{figure}[htb]
\begin{center}
\includegraphics[width=0.8\textwidth]{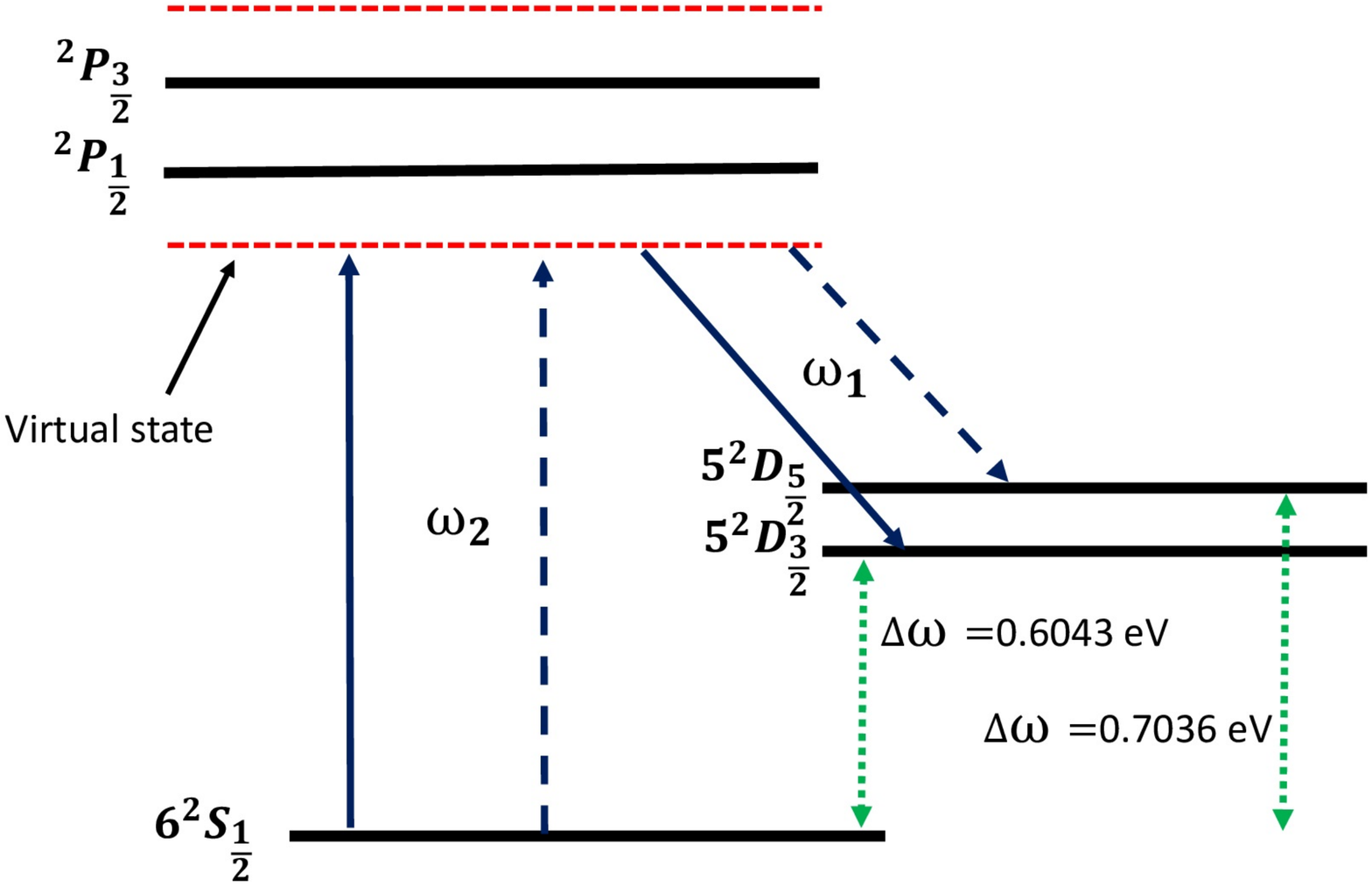}
\caption{Schematic energy level diagram for the ground and first few low-lying excited states (indicated by solid horizontal black line) of Ba$^+$ ion. The two-photon 6$^2S_\frac{1}{2}$ $-$ 5$^2D_\frac{3}{2}$ (indicated by solid indigo arrow) and 6$^2S_\frac{1}{2}$ $-$ 5$^2D_\frac{5}{2}$ (indicated by dotted indigo arrow) transitions are shown. The red dotted lines indicate the virtual states which can have different positions depending on the magic wavelengths. The green dotted lines with arrows in both directions indicate the frequencies of the clock transitions.}
\label{fig1}
\end{center}
\end{figure}

Two-photon spectroscopy is beneficial compared to single-photon spectroscopy in many aspects. A two-photon optical clock \cite{Hall1989,Martin2018,Martin2019,Perrella2019,Jackson2019,Gerginov2018} with a pair of counter-propagating laser beams is advantageous over a single-photon clock. Two-photon direct frequency comb spectroscopy (DFCS) enables detailed and precise studies of simultaneous investigation of the time-resolved atomic dynamic, spectral probing in the frequency domain, coherent accumulation and interference, and coherent control \cite{Marian2004, diddams2020}. Together, these photons will enable supreme control over the ions used as quantum gates or registers \cite{Kang2016}. Therefore, it can be a matter of significant research interest to explore how the dynamic polarizabilities and hence magic wavelengths for Ba$^+$ respond to a two-photon interaction. Especially, our interest is on minimizing the error budget in the frequency measurements on the optical clock transitions, 6$^2S_{\frac{1}{2}}$ $-$ 5$^2D_{\frac{3}{2}, \frac{5}{2}}$, of this ion due to the two-photon interaction. The three low-lying states of the Ba$^+$ ion form $\wedge$-shaped two-photon transitions between the 6$^2$S$_{\frac{1}{2}}$ and 5$^2$D$_{\frac{3}{2}, \frac{5}{2}}$ states, which is depicted in Fig.~\ref{fig1}.

In recent times, a number of highly accurate measurements were performed on the transition properties of Ba$^+$. Such measurements give us a way to select the necessary inputs for our study on the two-photon interaction process. As an example, we fix the difference of frequencies of the two counter-propagating laser lights from the measured transition frequencies of Ref.\cite{Dijck2015}. 
Moreover, the recent measurements \cite{Zhang2020} of branching fractions for the decays from the $^2P_{\frac{3}{2}}$ state give us a way to estimate the uncertainty in the dynamic polarizability values and hence the magic wavelengths \cite{Dutta2015}. 

We also demonstrate a distinct application of the two-photon dynamic polarizability in the spin-mixing processes \cite{Xu2012} of trapped atoms/ions inside spinor Bose-Einstein condensate (BEC). This type of condensates has been experimentally realized and theoretically investigated in different atom-atom \cite{Modgno2002, Thalhammer2008, McCarron2011, ARoy2015, Li2015, Mil2020, Chen2018, Li2020, Fang2020, Zhang2011, Li2017, Xu2010} and ion-atom mixtures \cite{Hirzler2020, Tomza2019}. For instance, the spin-oscillation is studied in an $^{87}$Rb--$^{23}$Na mixture \cite{Chen2018, Li2020} under the influence of a linearly polarized light. However, our curiosity is on two-photon induced spin-mixing processes in an ion-atom heteronuclear mixture, $^{137}$Ba$^+$--$^{87}$Rb. This mixture can be a potential candidate in ultra-cold chemistry and promising to reach the s-wave scattering regime with state-of-the-art experimental techniques \cite{Krych2011, Huber2014}. In general, in a binary species (X, Y) with the hyperfine quantum numbers ($F_X$, $F_Y$) = (1, 1), the spin-dependent inter-species interaction \cite{Widera2005} can induce different spin-mixing processes, such as $X_0+Y_0\Leftrightarrow X_{+1}+Y_{-1}$ and $X_0+Y_0\Leftrightarrow X_{-1}+Y_{+1}$, where the subscripts $0$,  $+1$, and $-1$ stand for the magnetic quantum numbers of the hyperfine levels. We recently showed that only one of these two processes could be achieved using a single linearly polarized focused vortex beam \cite{Bhowmik2020}.  However, the two-photon model can bring an interesting consequence in the controlling mechanism of both the spin-oscillation processes.

Our strategy to calculate the frequency-dependent or dynamic polarizability of an atomic state is described in detail in some of our earlier publications \cite{Dutta2015, Das2020}. The accuracy of the calculated polarizability value depends mainly on the exhaustiveness of the many-body methods used to evaluate the important electric dipole ($E$1) matrix elements. These matrix elements appeared in the sum-over-states expression of the polarizability \cite{Dutta2015, Das2020}. This work employs the highly accurate relativistic coupled-cluster method with single, double, and valence triple excitations (RCCSD(T)) \cite{Chaudhuri2003, Dutta2016, Dutta2013, Bartlett2007, Biswas2018,  Majumder2001, Bhowmik2017, Das2018} to compute the most important matrix elements \cite{Dutta2015, Das2020}. The rest of the important matrix elements are calculated using the second-order relativistic many-body perturbation theory (RMBPT(2)) \cite{Lindgren1986, Lindgren1985, Boyle1998, Johnson1996}. Section-\ref{TH} of this paper discusses a brief formalism to calculate a dynamic polarizability. In Section-\ref{RED}, we graphically depict and analyze our calculated dynamic polarizability values for Ba$^+$ considering both the single-photon and two-photon cases. Subsequently, we provide the two-photon magic wavelengths for the clock transitions. Following this, we illustrate the role of the two-photon polarizability in the controlling mechanism of the spin-oscillation processes for trapped $^{137}$Ba$^+$ $-$ $^{87}$Rb mixture. In Section-\ref{CON}, we conclude by highlighting the most important findings of this work along with their applications.

\section{THEORY} \label{TH}
The Stark-shift of an atomic state $|\psi_v\rangle$ in the presence of an external electric field is represented by using the second-order perturbation theory:
\begin{eqnarray}\label{1}
\Delta\xi_v(\omega)=-\frac{1}{2}\alpha_v(\omega)\mathcal{E}^{2}.
\end{eqnarray}
Here $\alpha_v(\omega)$ and $\mathcal{E}$ represent the dynamic polarizability and the applied field strength, respectively. `$v$' at the subscript indicates the single-valence atomic state having valence electron at the `$v$'th orbital. $\alpha_v(\omega)$ is comprised of three parts: valence part ($\alpha_v^V(\omega)$), core part ($\alpha_v^C(\omega$)), and core-valence part ($\alpha_v^{VC}(\omega)$) \cite{Bhowmik2018, Das2020}. Accordingly, one can write $\alpha_v(\omega)$=$\alpha_v^V(\omega)+\alpha_v^C(\omega)+\alpha_v^{VC}(\omega)$. Brief descriptions of all these three parts are given in Ref. \cite{Dutta2015}. The most crucial part in computing $\alpha_v(\omega)$ is $\alpha_v^V(\omega)$. The other two parts do not contribute significantly to the polarizability in general. Also, the core part of the polarizability is independent of any valence configuration. Therefore, in the determination of a magic wavelength, the values of $\alpha_v^C(\omega)$ or comprehensively $\alpha^C(\omega)$ are canceled mutually between the transition states. However, the values of $\alpha_v^{VC}(\omega$) are not the same for the transition states and hence can contribute by a tiny amount to a magic wavelength. Nevertheless, $\alpha_v^V(\omega)$ of the transition states play the determinant role in locating the position of a magic wavelength and thus are calculated with appropriate accuracy.

For a linearly polarized light, the valence part of the dynamic polarizability can be expressed in terms of scalar ($\alpha_v^{(s)}(\omega)$) and tensor ($\alpha_v^{(2)}(\omega)$) components as \cite{Dutta2015, Bhowmik2018, Das2020}
\begin{equation}\label{4}
[\alpha_v^V(\omega)]_{\text{LP}}=\underbrace{\frac{2}{3(2J_v + 1)}\sum_n d_{nv}}_{\text{\large$\alpha_v^{(s)}(\omega)$}}+C\times \underbrace{4\times \sqrt{\frac{5J_v(2J_v-1)}{6(J_v+1)(2J_v+1)(2J_v+3)}}\sum_n (-1)^{J_n+J_v} \left\{\begin{array}{ccc} J_v & 1 & J_n\\ 1 & J_v& 2 \end{array}\right \} d_{nv}}_{\text{\large$\alpha_v^{(2)}(\omega)$}},
\end{equation}
where \begin{equation}\label{69}
d_{nv}=\frac{|\langle\psi_v||d||\psi_n\rangle|^2 \omega_{nv}}{\omega_{nv}^2-\omega^2},\ \ \  C=\frac{3M_{J_v}^2-J_v(J_v+1)}{J_v(2J_v-1)}.
\end{equation}
Here `$n$' under the summation symbol indicates the different intermediate states. $\langle\psi_v||d||\psi_n\rangle$ is the reduced matrix element of the $E$1 operator between the atomic states $|\psi_v\rangle$ and $|\psi_n\rangle$. $\omega_{nv}=\epsilon_n-\epsilon_v$ is the excitation energy between these two states. Including the contributions of $\alpha^C(\omega)$ and $\alpha^{VC}_{v}(\omega)$, the total scalar polarizability becomes $\alpha_v^{(0)}(\omega)$=$\alpha_v^{(s)}(\omega)$+$\alpha^C(\omega)$+$\alpha_v^{VC}(\omega)$.  The BBR shift between the transition states can be estimated from the differential Stark-shift for the scalar polarizabilities of the states at $\omega=0$ \cite{Porsev2006, Mitroy2010}. 
 
The vector component ($\alpha^{(1)}_v(\omega)$) appears in the valence part of the dynamic polarizability when a circularly polarized light is considered. Adding this component, the resultant expression for the valence polarizability becomes \cite{Das2020}
\begin{equation}
\alpha_v^V(\omega)=[\alpha^{V}_v(\omega)]_{\text{LP}}+A\frac{M_{J_v}}{2J_v}\alpha^{(1)}_v(\omega),
\label{eq5}
\end{equation}
with
\begin{equation}\label{11}
\alpha^{(1)}_v(\omega)=-\sqrt{\frac{6J_v}{(J_v+1)(2J_v+1)}}\sum_n (-1)^{J_n+J_v} \left\{\begin{array}{ccc} J_v & 1 & J_v\\ 1 & J_n& 1 \end{array}\right \} 
\left(\frac{2\omega}{\omega_{nv}}\right)d_{nv}.
\end{equation}
The value of $A$ is 0 for a linearly polarized light, +1 for a left circularly polarized light, and $-$1 for a right circularly polarized light. 
  
In the presence of two counter-propagating laser beams with frequency $\omega_1$ and $\omega_2$, the Stark-shift of the atomic state $|\psi_v\rangle$ can be represented by
\begin{equation}\label{99}
\Delta\xi_v(\omega)= \frac{1}{2}\mathcal{E}_1^{2}[\alpha_{v}(\omega_1)+ \alpha_{v}(\omega_{2})\frac{\mathcal{E}_2^{2}}{\mathcal{E}_1^{2}}]=\frac{1}{2}\mathcal{E}_1^{2}\alpha_{tp}(\omega_1, \omega_2).
\end{equation}
Here we assume that the two fields of the laser beams with intensities $\mathcal{E}_1^{2}$ and $\mathcal{E}_2^{2}$, respectively, act on the atom or ion independently.  Here $\alpha_{tp}(\omega_1, \omega_2)=\alpha_{v}(\omega_1)+ \alpha_{v}(\omega_{2})\frac{\mathcal{E}_2^{2}}{\mathcal{E}_1^{2}}$ is defined as the two-photon polarizability.

\section{Results and Discussions}\label{RED}

\subsection{Parameters used in our calculations for Ba$^+$}
The precursor of employing any correlated many-body method to an atomic system is to construct accurate Dirac-Fock (DF) orbital wavefunctions. In this work, we use Gaussian type orbital (GTO) basis functions \cite{Clementi1990,Dutta2016} with optimized \cite{Roy2015} even-tempered exponents \cite{Huzinaga1993, Huzinaga1985} to generate the DF wavefunctions. These exponents, optimized for all the relativistic symmetries, are presented in Table~\ref{xxx}. The number of GTO basis functions used in our calculations for the $s$-, $p$-, $d$-, $f$-, $g$-, and $h$-type symmetries are 33, 30, 28, 25, 21, and 20 respectively. The numbers of active DF orbitals, chosen for our correlated many-body calculations using the RMBPT(2) and the RCCSD(T) methods, are considered as 12, 12, 13, 10, 6, and 6 respectively, for the above-mentioned orbital symmetries.
\begin{table}[htb]
\centering
\caption{Even-tempered exponents \cite{Huzinaga1985} used in our relativistic calculations. Ref.\cite{Huzinaga1985} describe these exponents for non-relativistic symmetries, whereas we use these for relativistic symmetries.}
\begin{tabular}{c c c c c c c c c c c c}
\hline\hline
Symmetry & $s_{1/2}$  &  $p_{1/2}$   & $p_{3/2}$  & $d_{3/2}$ & $d_{5/2}$ &  $f_{5/2}$ \\
 $m$  &  1 &  2  & 3  & 4   & 5  & 6 \\ 
\{$\alpha_m$, $\beta_m$\} & \{0.0020, 2.80\} & \{0.0060, 2.55\}  & \{0.0045, 1.58\} & \{0.0050, 2.20\} & \{0.0055, 2.00\} & \{0.0087, 2.65\} \\
\hline
Symmetry & $f_{7/2}$  & $g_{7/2}$   & $g_{9/2}$  & $h_{9/2}$ & $h_{11/2}$\\
 $m$  & 7 & 8 & 9 & 10 & 11\\ 
\{$\alpha_m$, $\beta_m$\} & \{0.0033, 2.45\} & \{0.0077, 2.73\} & \{0.0077, 2.73\} & \{0.0063, 2.73\} & \{0.0063, 2.73\} \\
\hline
\hline
	\end{tabular}
	\label{xxx}
\end{table}

\subsection{Electric dipole reduced matrix elements of Ba$^+$}
In Table \ref{2}, we present the absolute values of a few $E$1 reduced matrix elements computed using the RCCSD(T) method and compare these values with some earlier results. All these reduced matrix elements can be crucial in computing the dynamic polarizabilities of 6$^2S_{\frac{1}{2}}$, 5$^2D_{\frac{3}{2}}$, and 5$^2D_{\frac{5}{2}}$ states, which is indicated in the next subsection. The SD and SDpT values are calculated by Safronova \cite{Safronova2010} and Barrett {\it et al.} \cite{Barrett2019}. The SD method is a linearized approximation of the coupled-cluster method with single and double excitations, whereas SDpT is an extension of the SD method to include partial triple excitations. Safronova did not present the SD and SDpT reduced matrix elements associated with the n$^2F_{\frac{5}{2}, \frac{7}{2}}$  states in her paper due to some convergence issue \cite{Safronova2010}. Therefore, she presented the third-order relativistic perturbation theory results ($Z^{DF+2+3}_{vw}$) for the matrix elements associated with these states. This convergence issue was resolved later in the work of Barrett \textit{et al.} \cite{Barrett2019}.  Sahoo \textit{et al.} \cite{Sahoo2009} employed the similar RCCSD(T) method as used by us. But they used different sets of basis parameters and active orbitals. The dipole matrix elements for Ba$^{+}$ were calculated by Jian {\it et al.} using the relativistic configuration interaction plus core polarization (RCICP) method \cite{Jiang2021}. They used these matrix elements to compute dynamic polarizabilities and magic wavelengths for this ion due to single-photon interaction. All these accurate theoretical values and some experimental results as found in the literature can give a good understanding of the precision of our calculated polarizability values and hence magic wavelengths. The agreements between our values and the other values are excellent for the $E$1 matrix elements associated with the lower states. However, the $E$1 matrix elements corresponding to some of the higher states agree poorly. Even mutual agreements among these latter matrix elements, calculated by using different theoretical approaches, are poor. Fortunately, such matrix elements have little contributions to the dynamic polarizabilities. Therefore, the accuracy of these matrix elements hardly has any impact on our estimated magic wavelengths which are discussed later.
\begin{table} [htb]
\centering
\caption{The absolute values of our calculated reduced electric dipole matrix elements (indicated by `Our') in a.u. are compared with few other theoretical and experimental (Expt.) values.}
\begin{tabular}{c c c c c c c}
\hline\hline
Transition & RCCSD(T) \cite{Sahoo2009} & RCICP \cite{Jiang2021} & SD \cite{Safronova2010} & SDpT \cite{Safronova2010} & Our & Expt.\\
\hline
		5 $^2D_\frac{3}{2}$ $\rightarrow$ 6 $^2P_\frac{1}{2}$ & 3.11(3)&3.033(29)&3.0503&3.0957&3.0641&3.03(9)$^{a}$, 2.90(9)$^{b}$, 3.034$^{c}$\\
		$\rightarrow$6 $^2P_\frac{3}{2}$ &1.34(2)&1.336(13)&1.3324&1.3532&1.3335&1.36(4)$^{a}$, 1.54(19)$^{b}$, 1.325$^{c}$\\
		&&&&&&1.349(36),1.33199(96)$^{e}$\\
		$\rightarrow$7 $^2P_\frac{1}{2}$& 0.28(2)&0.23(3)& 0.2792& 0.2775& 0.3336&0.42(11)$^{b}$\\
		$\rightarrow$7 $^2P_\frac{3}{2}$& 0.16(1)&0.14(2)&0.1555& 0.1548&0.1797&0.19(5)$^{b}$\\
		$\rightarrow$8 $^2P_\frac{1}{2}$& 0.07(2)&0.10(5)&0.1346& 0.1349&0.1527&0.23(6)$^{b}$\\
		$\rightarrow$8 $^2P_\frac{3}{2}$&0.07(2)&0.07(3)&0.0768& 0.0769& 0.0867& 0.10(3)$^{b}$\\
		5 $^2D_\frac{5}{2}$ $\rightarrow$ 6 $^2P_\frac{3}{2}$ & 4.02(7)&4.105(39)&4.1032& 4.1631&4.1368& 4.080$^{c}$, 4.1028(25)$^{e}$\\
		7 $^2P_\frac{3}{2}$&0.46(1)&0.39(6)&0.4513& 0.4500& 0.5123&\\
		8 $^2P_\frac{3}{2}$&0.21(2)&0.18(3)&0.2232& 0.2239& 0.2457&\\
		6 $^2S_\frac{1}{2}$ $\rightarrow$ 6 $^2P_\frac{1}{2}$ & 3.36(1)&3.275(47)& 3.3380 & 3.3710& 3.4082& 3.36(16)$^{a}$, 3.36(4)$^{b}$, 3.3357$^{c}$\\
		$\rightarrow$ 6 $^2P_\frac{3}{2}$ & 4.73(3)&4.637(67)& 4.7097& 4.7569& 4.8103& 4.45(19)$^{a}$, 4.55(10)$^{b}$, 4.72(4)$^{d}$ \\
		&&&&&&4.7065$^{c}$\\
		$\rightarrow$ 7 $^2P_\frac{1}{2}$ & 0.10(1)&0.10(5)& 0.0605& 0.0607& 0.0350& 0.24(3)$^{b}$ \\
		$\rightarrow$ 7 $^2P_\frac{3}{2}$ &0.17(5)&0.04(2)& 0.0870& 0.0858& 0.1337& 0.33(4)$^{b}$ \\
		$\rightarrow$ 8 $^2P_\frac{1}{2}$ & 0.11(5)&0.11(6)& 0.0868& 0.0866&0.0426& 0.10(1)$^{b}$ \\
		$\rightarrow$ 8 $^2P_\frac{3}{2}$ & 0.11(5)&0.06(3)& 0.0331&0.0334&0.0400&0.15(2)$^{b}$\\
		\hline
		Transition & RCCSD(T)\cite{Sahoo2009} &RCICP \cite{Jiang2021} & SD\cite{Barrett2019} & SDpT \cite{Barrett2019}& Our & Z$_{\textit{\textit{vw}}}^{DF+2+3}$\cite{Safronova2010}\\
		\hline
		5 $^2D_\frac{3}{2}$ $\rightarrow $4 $^2F_\frac{5}{2}$&3.75(11)&3.671(35)&-&-&3.6370&3.6216\\
		$\rightarrow$5 $^2F_\frac{5}{2}$&1.59(8)&-&-&-&1.9454&1.8513\\
		$\rightarrow$6 $^2F_\frac{5}{2}$&0.17(2)&-&-&-&1.15358&0.9208\\
		5 $^2D_\frac{5}{2}$ $\rightarrow $4 $^2F_\frac{5}{2}$&1.08(4)&1.002(9)&0.998& 1.012&1.0044& 0.9951\\
		$\rightarrow $4 $^2F_\frac{7}{2}$&4.84(5)&4.500(42)&4.475& 4.540&4.5017&4.4504\\
		$\rightarrow $5 $^2F_\frac{5}{2}$&0.45(7)&-&0.016& 0.210&0.5255&0.5005\\
		$\rightarrow $5 $^2F_\frac{7}{2}$&2.47(6)&-&0.130& 1.049&2.4283&2.2445\\
		$\rightarrow $6 $^2F_\frac{5}{2}$&0.15(2)&-&0.236& 0.018&0.2996&0.2449\\
		$\rightarrow $6 $^2F_\frac{7}{2}$&1.04(7)&-&0.961& 0.170&1.3690&1.1160\\
		\hline
	\end{tabular}
	\label{2}
	$a\rightarrow$ Ref.\cite{Kastberg1993}, $b\rightarrow$ Ref.\cite{Klose2002, Davidson1992},$c\rightarrow$ Ref.\cite{Woods2010}, $d\rightarrow$ Ref.\cite{Kurz2008}, $e\rightarrow$ Ref.\cite{Zhang2020}
\end{table}

\subsection{Frequency-dependent or dynamic dipole polarizabilities of Ba$^+$}
As mentioned in the Section-\ref{TH}, the calculation of the polarizability can be divided into three parts:  $\alpha^{C}$, $\alpha_v^{VC}$, and $\alpha_v^{V}$. Our calculation strategy for these parts for an alkali-metal-like system is described in our very recent work Ref.\cite{Das2020}. $\alpha^{C}$, which is calculated by using the third-order relativistic many-body perturbation theory (RMBPT(3)) \cite{Dutta2015, Das2020, Dutta2020}, varies very slowly with the frequency within the frequency region considered here. The static value of the core polarizability ($\alpha^C(\omega=0)$) is calculated to be 10.79 a.u.. The dynamic core polarizability becomes maximum with a value of 11.10 a.u. at the highest frequency ($\omega$ = 0.18 a.u.) in this region. $\alpha_v^{VC}$ is almost insensitive to the frequency of the external field. Its values are calculated to be -0.48 a.u., -0.69 a.u. and -0.82 a.u. for the 6$^2S_\frac{1}{2}$, 5$^2D_{\frac{3}{2}}$ and 5$^2D_{\frac{5}{2}}$ states, respectively, and are kept fixed within the entire frequency region. However, $\alpha^{V}_{v}$, which contributes mostly to the polarizability, is very much sensitive to the frequency in this region. Therefore, calculation of this part for the 6$^2S_\frac{1}{2}$, 5$^2D_{\frac{3}{2}}$, and 5$^2D_{\frac{5}{2}}$ states should be performed with utmost care to determine the frequency dependence of the polarizability values accurately. The most dominant contributions to the $\alpha_v^{V}(\omega)$ appear from the sum of the terms having matrix elements associated with the intermediate states 6--9$^2P_{\frac{1}{2},\frac{3}{2}}$ and 4--6$^2F_{\frac{5}{2},\frac{7}{2}}$. Here we use the RCCSD(T) method to calculate these matrix elements. The next significant contributions arise from the intermediate states 10--15$^2P_{\frac{1}{2},\frac{3}{2}}$ and 7--15$^2F_{\frac{5}{2},\frac{7}{2}}$, where the matrix elements are calculated using the RMBPT(2) method. The excitation energies associated with all these low-lying states are extracted from the website of NIST atomic database \cite{NIST2020}. The remaining contributions, which are relatively small, include the sum of the terms having intermediate states 16--25$^2P_{\frac{1}{2},\frac{3}{2}}$ and 16--25$^2F_{\frac{5}{2},\frac{7}{2}}$, and are calculated using the DF method. We do not consider any intermediate state beyond 25$^2P_{\frac{1}{2}, \frac{3}{2}}$ and 25$^2F_{\frac{5}{2}, \frac{7}{2}}$ due to almost negligible contributions from these.

\subsection{Differential scalar polarizabilities of Ba$^+$}

\begin{figure}[htb]
\begin{center}
\includegraphics[width=0.6\textwidth]{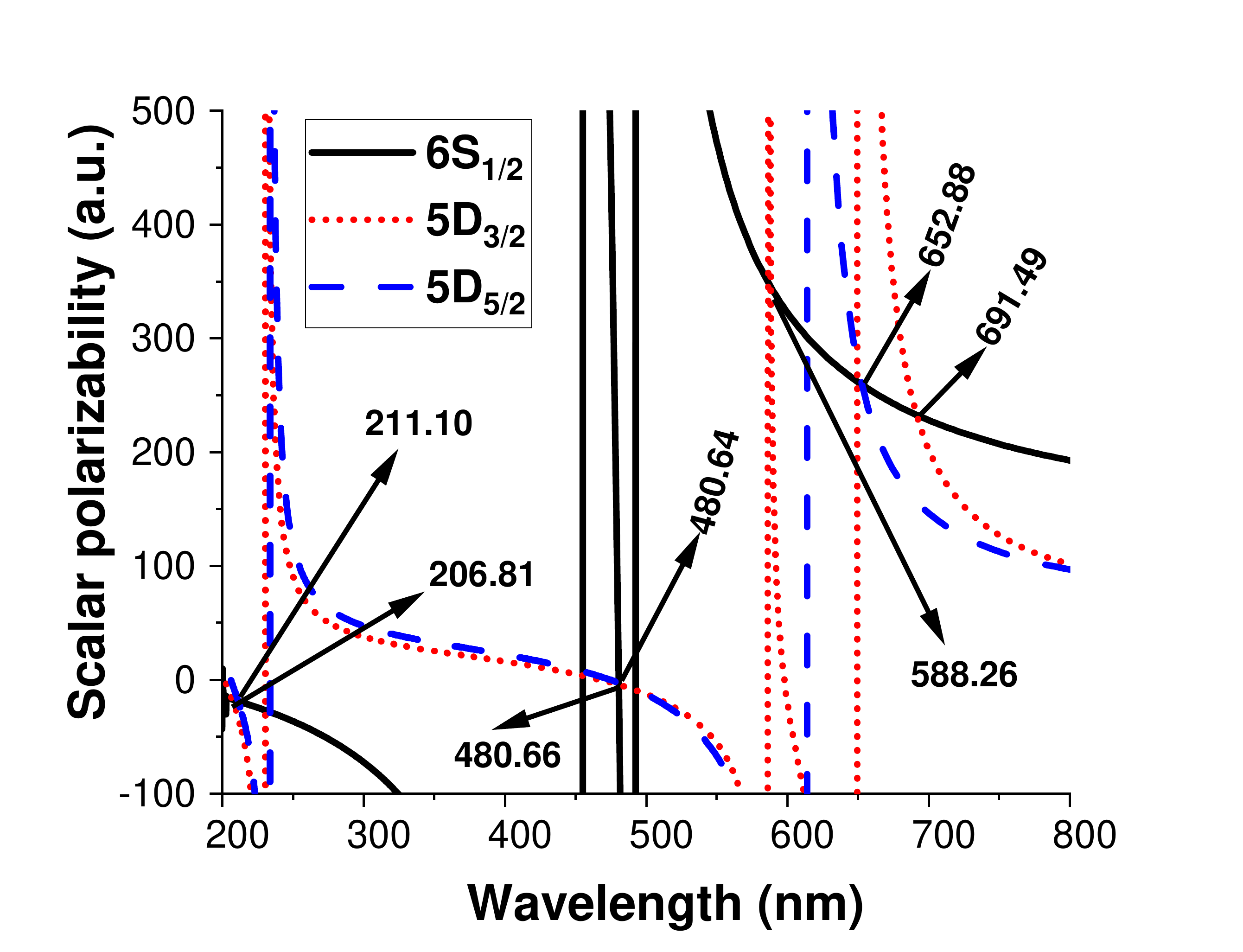}
\caption{Dynamic scalar polarizabilities of 6$^2S_\frac{1}{2}$, 5$^2D_\frac{3}{2}$, and 5$^2D_\frac{5}{2}$ states of Ba$^+$. Here the different n$^{2}L_{J_v}$ states are indicated by nL$_{J_v}$. The crossing points between the polarizability curves for 6$^2S_\frac{1}{2}$ and 5$^2D_{\frac{3}{2}, \frac{5}{2}}$ states provide the values of wavelengths (indicated in the panels in the unit of nm) which yield zero differential scalar polarizabilities for the 6$^2S_\frac{1}{2}$ $-$ 5$^2D_{\frac{3}{2}, \frac{5}{2}}$ clock transitions.}
\label{fig2}
\end{center}
\end{figure}

As a preliminary test of the quality of our calculations aiming at magic wavelengths, we compare our differential scalar polarizability values for the clock transitions with the corresponding other values as available in the literature. Fig. \ref{fig2} displays the variations of the dynamic scalar polarizabilities ($\alpha_v^{(0)}(\omega)$) with wavelength for 6$^2S_\frac{1}{2}$, 5$^2D_\frac{3}{2}$ and 5$^2D_\frac{5}{2}$ states. The crossing points between the polarizability curves of 6$^2S_\frac{1}{2}$ and 5$^2D_{\frac{3}{2}, \frac{5}{2}}$ states indicate the zero differential scalar polarizabilities ($\Delta \alpha^{(0)}(\omega)$=0) for the 6$^2S_\frac{1}{2}$ -- 5$^2D_{\frac{3}{2}, \frac{5}{2}}$ transitions. In Table \ref{1}, we report the differential static scalar polarizabilities ($\Delta \alpha^{(0)}(0)$) which can provide estimations of the BBR shifts in the clock transitions. These $\Delta \alpha^{(0)}(0)$ values are compared with the corresponding recent estimations of Barrett {\it et al.} \cite{Barrett2019}, Chanu {\it et al.} \cite{Chanu2020} and Sahoo {\it et al.} \cite{Saho2009}. This comparison shows a good agreement between our values and their values. This table also contains the positions of the crossing points (wavelengths indicated by $\lambda_0$) as mentioned above. Barrett {\it et al.} and  Chanu {\it et al.} reported only one such crossing point for the 6$^2S_\frac{1}{2}$ -- 5$^2D_{\frac{5}{2}}$ clock transition, which is located near 653 nm. This result matches excellently with our estimation of 652.88 nm. However, the table contains a few additional values of $\lambda_0$ for this clock transition and also, the $\lambda_0$ values for the 6$^2S_\frac{1}{2}$ -- 5$^2D_\frac{3}{2}$ clock transition. 

\begin{table} [htb]
\centering
\caption{Comparisons of our (`Our') calculated differential static scalar polarizabilities, i.e., $\Delta\alpha^{(0)}(0)$ for the clock transitions with the corresponding other (`Other') values as found in the literature. The last two columns highlight the comparison of $\lambda_0$ at which $\Delta\alpha^{(0)}(\omega) =0$.}
	
	\begin{tabular}{c c r c l c }
		
		\hline\hline
		Clock transition &\multicolumn{2}{c}{$\Delta\alpha^{(0)}(0)$ (a.u.)}&\ &\multicolumn{2}{c}{$\lambda_0$ (nm)}\\
		&Our&Other&\ & Our&Other\\
		\hline
		6$^2S_\frac{1}{2}$--5$^2D_\frac{5}{2}$& -71.39& -73.1(1.3) \cite{Barrett2019}, & &652.88, & 653(near) \cite{Barrett2019,Chanu2020}\\
		&& -73.56(21) \cite{Chanu2020},&&480.64, 211.10& -\\
		&&-73.59 \cite{Saho2009} &&&\\
		6$^2S_\frac{1}{2}$--5$^2D_\frac{3}{2}$&-77.64&-75.45 \cite{Saho2009}&&691.49, 588.26& -\\
		& & &&480.66, 206.81&-\\
		\hline
	\end{tabular}
	\label{1}
\end{table}

\subsection{Single-photon dynamic polarizabilities and magic wavelengths of Ba$^+$}
\begin{figure}[htb]
\subfloat{\includegraphics[width=0.52\linewidth]{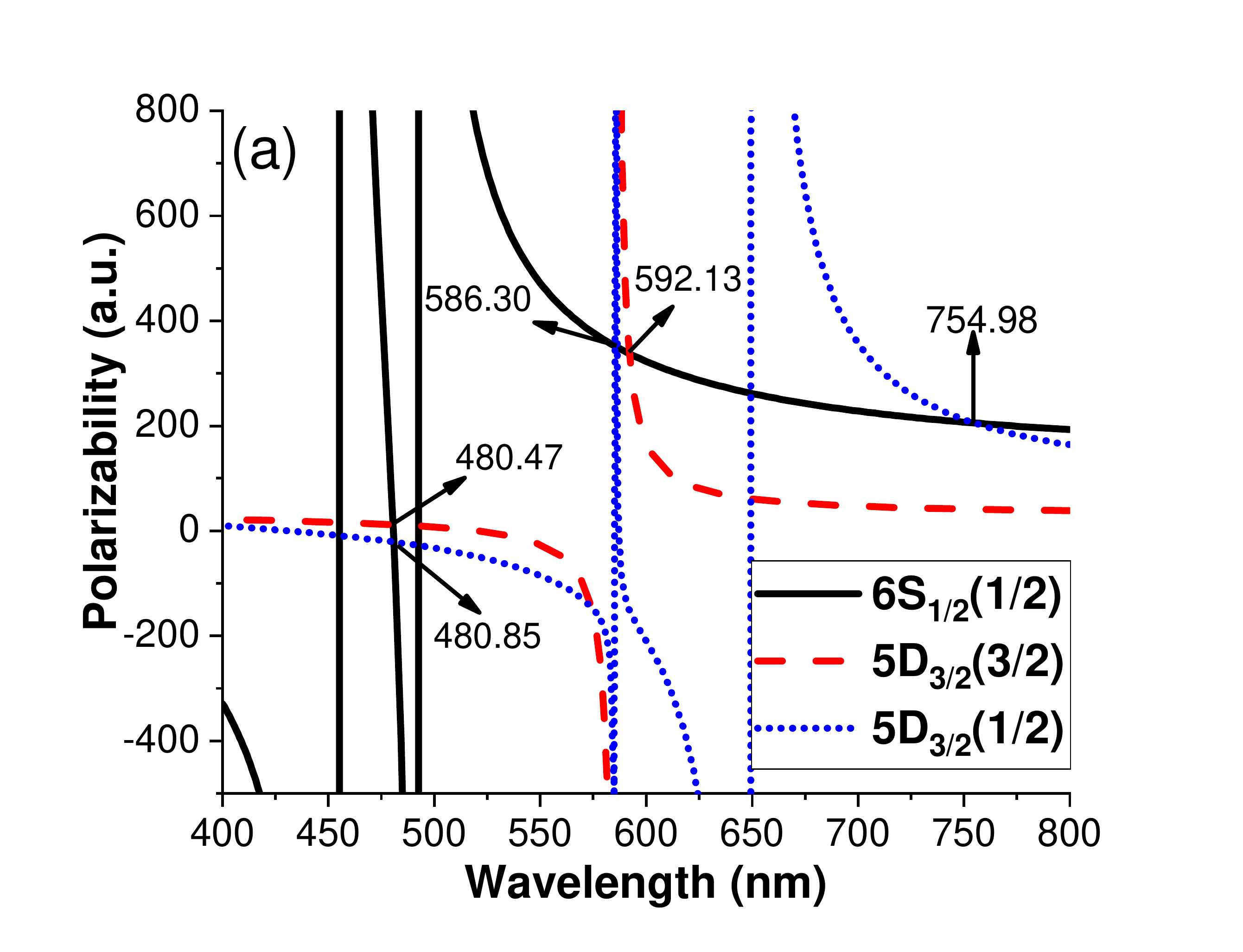}}
\subfloat{\includegraphics[width=0.52\linewidth]{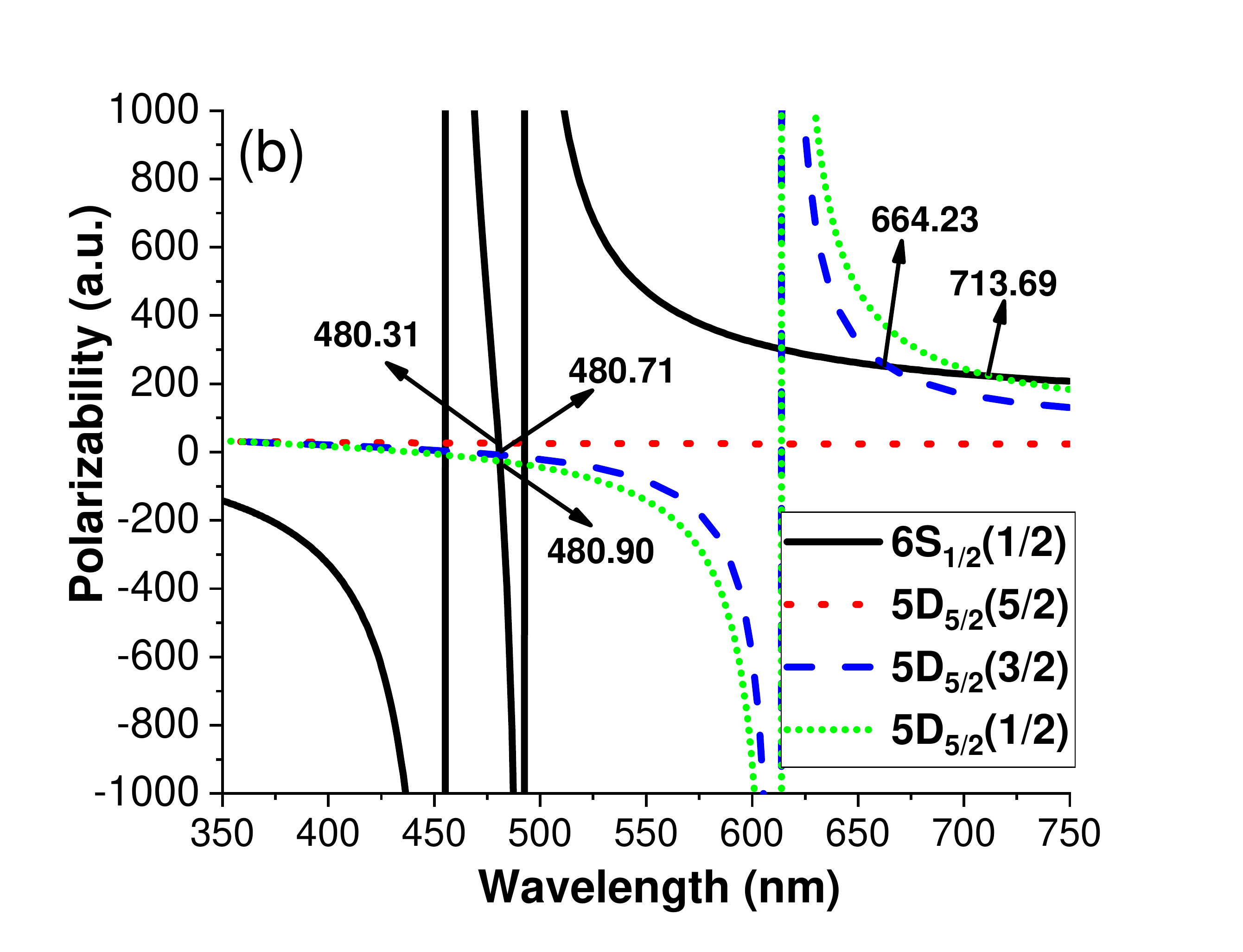}}
\caption{Variations of the polarizabilities with wavelength for the 6$^2S_\frac{1}{2}$ and 5$^2D_{\frac{3}{2}, \frac{5}{2}}$ states to extract the magic wavelengths (indicated in the panels in the unit of nm) in the presence of a linearly polarized light source. Here the different $(J_v, M_{J_v})$  levels of the state n$^{2}L_{J_v}$ are indicated by nL$_{J_v}(M_{J_v})$.}
\label{fig3}
\end{figure}

Fig. \ref{fig3} represents the dynamic polarizability profiles for the 6$^2S_\frac{1}{2}$ ($M_{J_v}$=$\frac{1}{2}$), 5$^2D_\frac{3}{2}$($M_{J_v}$=$\frac{3}{2}$,$\frac{1}{2}$), and 5$^2D_\frac{5}{2}$($M_{J_v}$=$\frac{5}{2}$,$\frac{3}{2}$,$\frac{1}{2}$) states in the presence of a single linearly polarized laser light. The crossing points between the two profiles for the 6$^2S_\frac{1}{2}$ and 5$^2D_{\frac{3}{2}, \frac{5}{2}}$ states indicate the magic wavelengths for the transitions between them. The magic wavelengths (in nm) in the optical region are highlighted in the figure and also tabulated in Table \ref{3}. We do not highlight and tabulate any magic wavelength which is obtained on a sharp (vertical) line. The magic wavelengths on these lines are of no practical use. The magic wavelengths in the optical region can be favoured due to the easy availability of lasers operated in this region. However, the magic wavelengths, which occur with high polarizability values, can be the candidates of optimal choice. Table \ref{3} shows that our calculated magic wavelengths agree excellently on average with the corresponding magic wavelengths reported in Ref.\cite{Jiang2021} and Ref.\cite{Kaur2015}. This agreement can indicate the reliability of our computed two-photon magic wavelengths which are described in the next subsection.

\begin{table}[htb]
\centering
\caption{Single-photon magic wavelengths ($\lambda_{m}$) in nm and the corresponding polarizabilities ($\alpha$) in a.u. for the clock transitions 6$^2S_\frac{1}{2}(\frac{1}{2})$ -- 5$^2D_{J_v}(M_{J_v})$ are presented. Our estimated magic wavelengths (`Our') are also compared with the corresponding values in Ref. \cite{Kaur2015} and Ref. \cite{Jiang2021}.}
\begin{tabular}{c c c c c }
		\hline\hline
		States & \multicolumn{3}{c}{$\lambda_{m}$} & $\alpha$ \\
		(${J_v}$,$M_{J_v}$) & Our & Ref. \cite{Kaur2015} & Ref. \cite{Jiang2021} & \\
		\hline	
		($\frac{3}{2},\frac{3}{2}$)&592.13&592.46&592.39(14)&337.27
		\\
		&480.47&480.44&480.539(14)& 12.03\\		
		($\frac{3}{2},\frac{1}{2}$)&754.98&767.81&757.7(3.9)&337.27 \\
		&586.30&585.98&585.982(10)&348.92 \\
		&480.85&480.81&480.93(3)&-21.13 \\
		($\frac{5}{2},\frac{5}{2}$)&480.31&480.26&480.38(2)&26.3 
		\\
		($\frac{5}{2},\frac{3}{2}$)&664.23&666.64&663.6(1.4)&250.17 
		\\
		&480.71&480.71&480.86(2)&-8.58
		\\	
		($\frac{5}{2},\frac{1}{2}$)&713.69&718.18&707.9(3.3)&221.45	
		\\
		&480.90&480.93&481.10(2)&-26.17 \\	
		\hline
	\end{tabular}
	\label{3}
\end{table}

\subsection{Two-photon dynamic polarizabilities and magic wavelengths of Ba$^+$}  
\begin{figure}[H]
\centering
\subfloat{\includegraphics[width=0.52\linewidth]{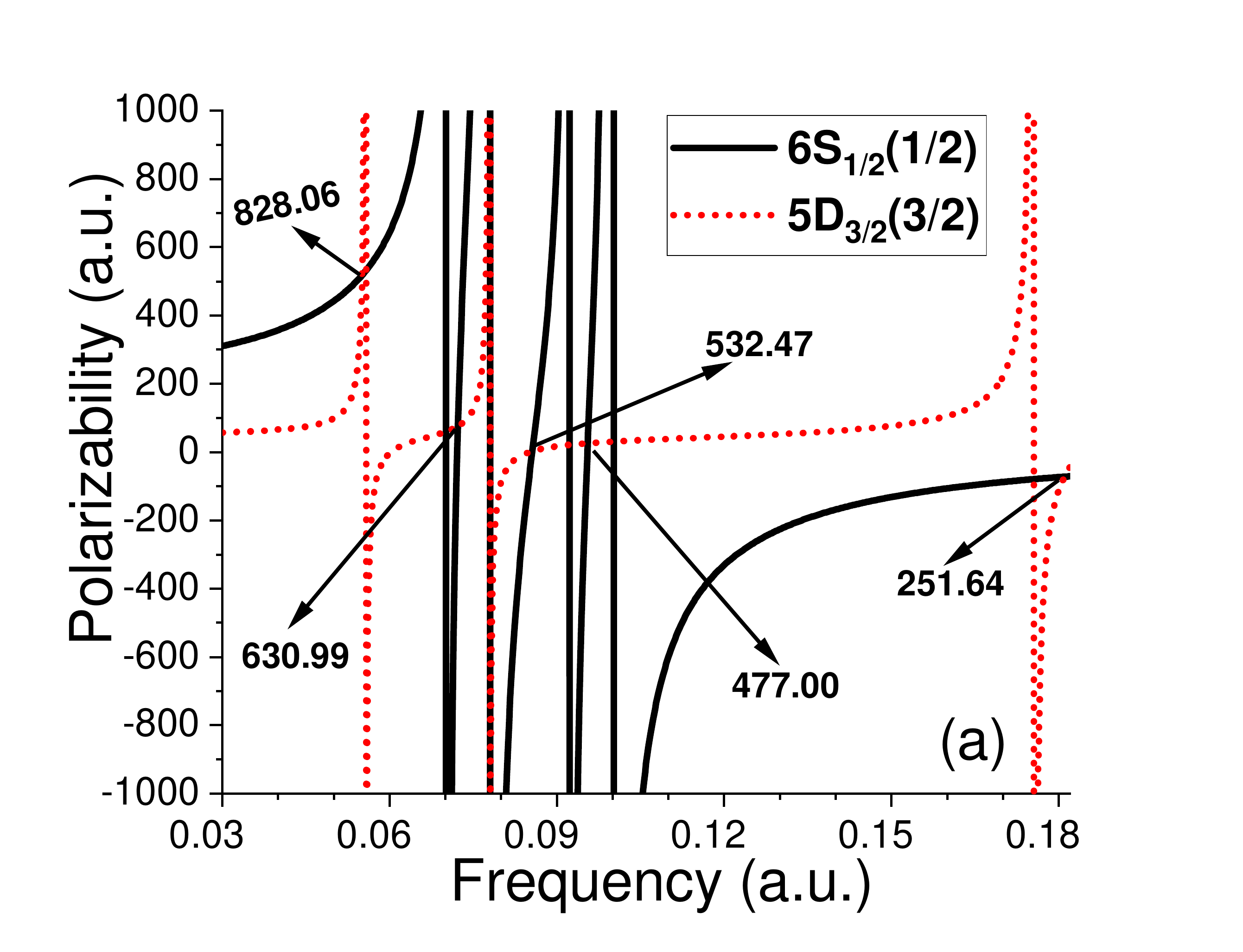}}
\subfloat{\includegraphics[width=0.52\linewidth]{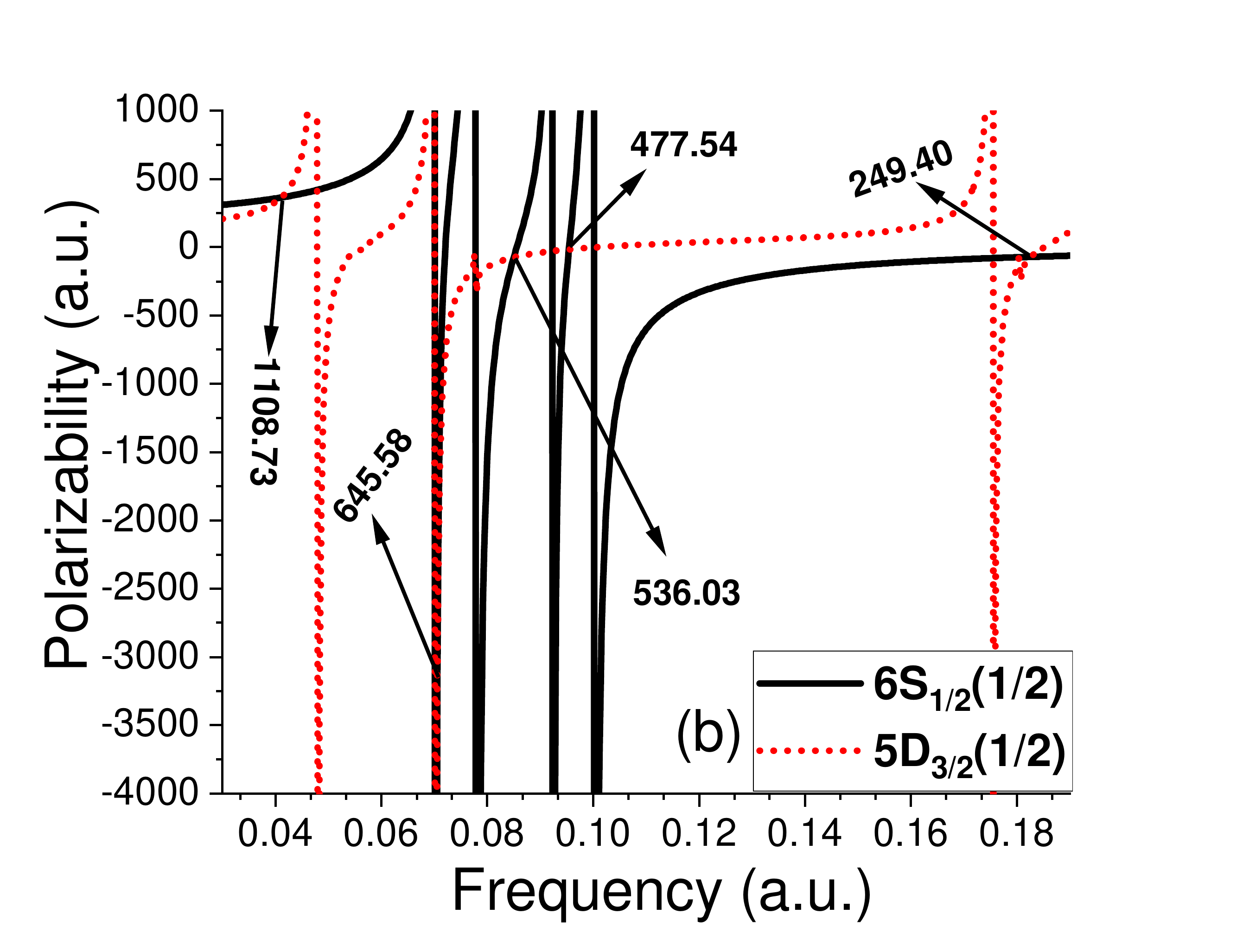}}\\
\hspace{2cm}6$^2S_\frac{1}{2}$ $-$ 5$^2D_\frac{3}{2}$ two-photon transition
\subfloat{\includegraphics[width=0.52\linewidth]{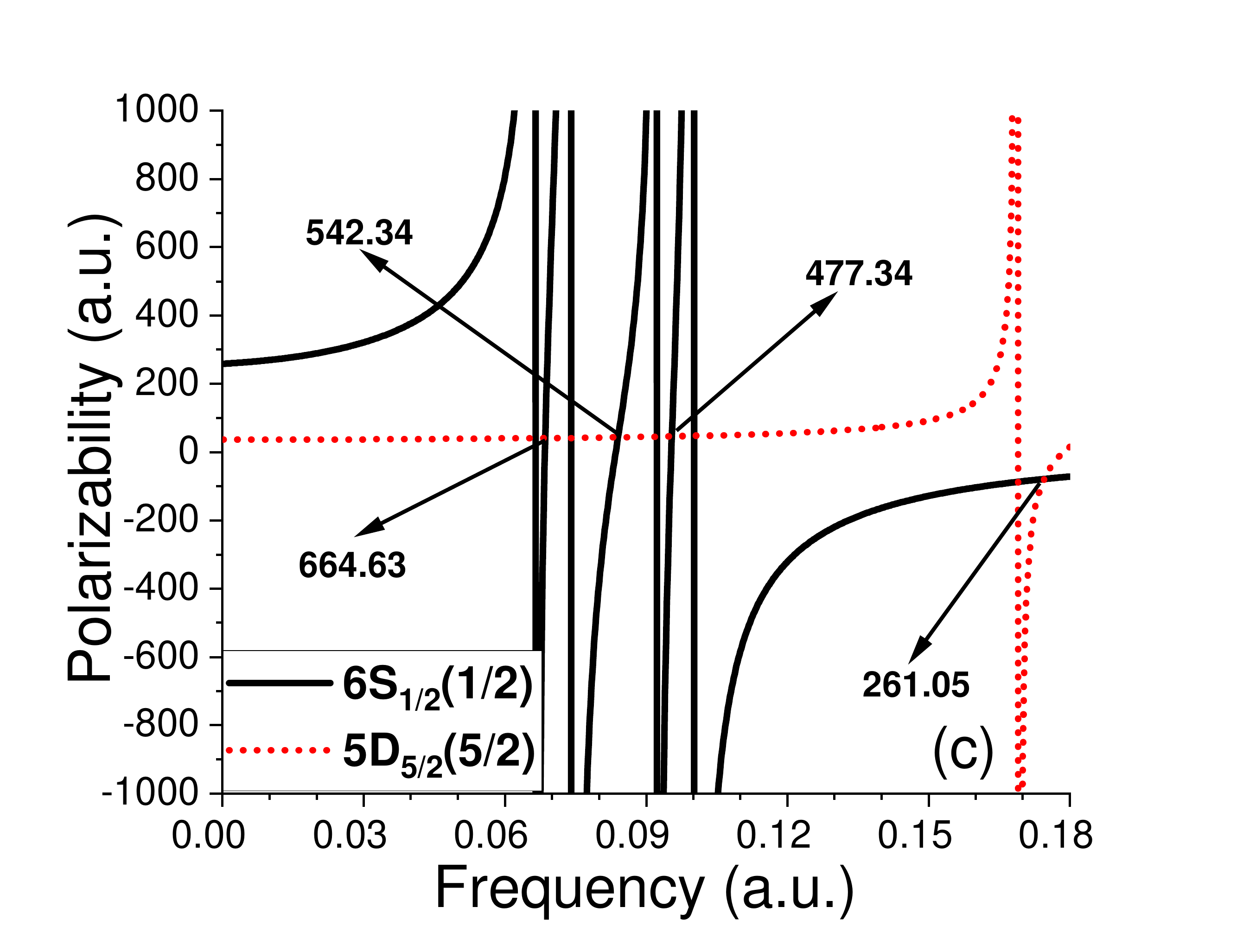}}
\subfloat{\includegraphics[width=0.52\linewidth]{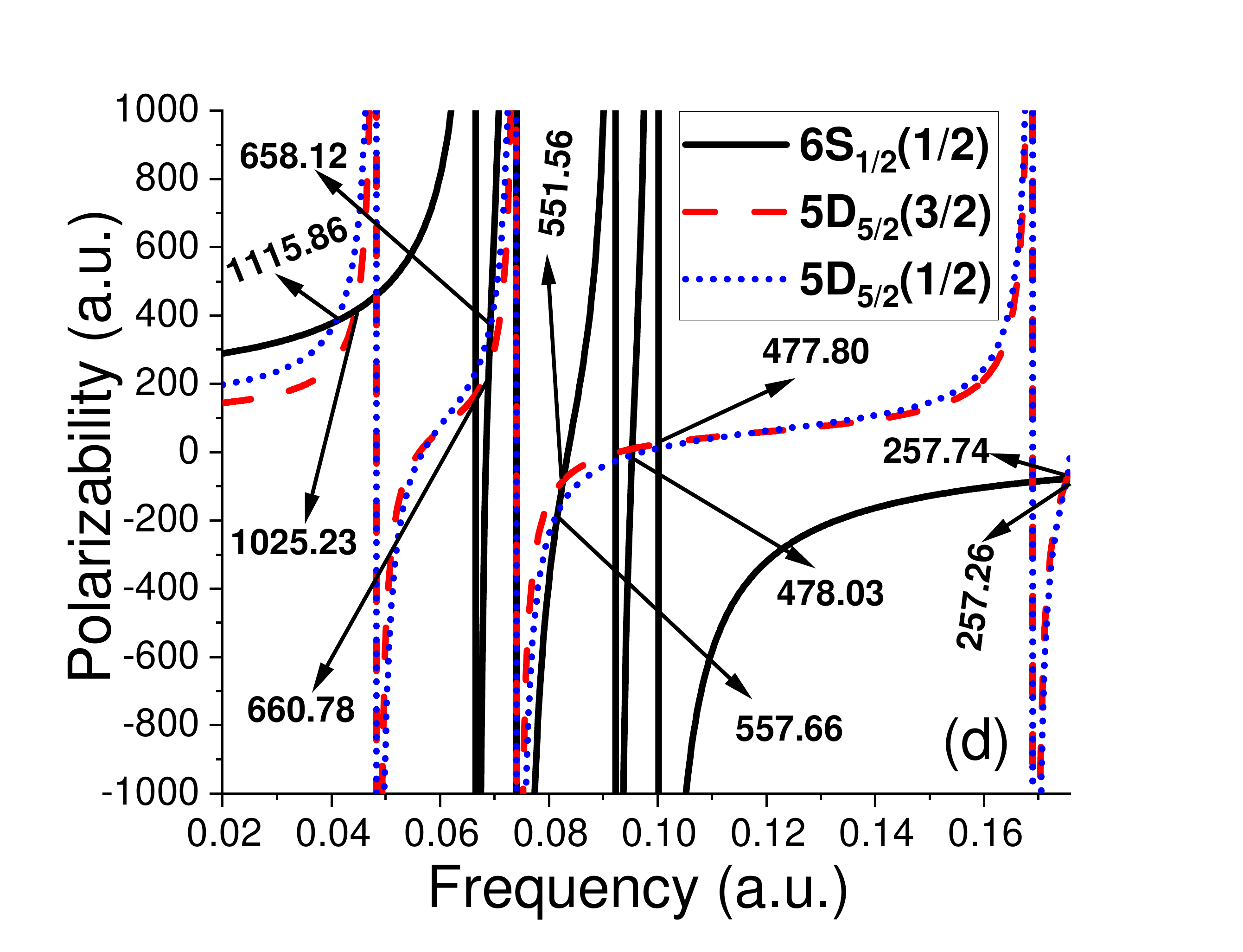}}\\
\hspace{2cm}6$^2S_\frac{1}{2}$ $-$ 5$^2D_\frac{5}{2}$ two-photon transition
\caption{Variations of the polarizabilities with frequency for the 6$^2S_\frac{1}{2}$ and 5$^2D_{\frac{3}{2}, \frac{5}{2}}$ states to extract the magic wavelengths (indicated in the panels in the unit of nm) in the presence of two counter-propagating linearly polarized lights. Here the different $(J_v, M_{J_v})$  levels of the state n$^{2}L_{J_v}$ are indicated by nL$_{J_v}(M_{J_v})$.}
\label{fig4}
\end{figure}

The primary objective of the present work is to estimate the two-photon magic wavelengths accurately. The estimated values of these wavelengths are tabulated in Table \ref{9} for the 6$^2S_\frac{1}{2}$ -- 5$^2D_{\frac{3}{2},\frac{5}{2}}$ clock transitions. Here we assume that both the counter-propagating laser beams have linear polarizations. We fix the difference ($\Delta \omega$ = $\omega_2-\omega_1$) between the frequencies of the two beams at the frequency of the corresponding clock transition. Accordingly, the value of $\Delta \omega$ is 0.022206903 a.u. \cite{Dijck2015} for the 6$^2S_\frac{1}{2}$--5$^2D_\frac{3}{2}$ clock transition and 0.025856323 a.u. \cite{NIST2020} for the 6$^2S_\frac{1}{2}$--5$^2D_\frac{5}{2}$ clock transition. The intensities of the two laser beams at the position of the Ba$^+$ ion are considered to be the same ($\mathcal{E}_1^2=\mathcal{E}_2^2$). Therefore, following Eq.~(\ref{99}), the effective or two-photon polarizability of a state can be written as the sum of the independent polarizabilities caused by each of the two laser beams: $\alpha_{tp}(\omega_1, \omega_2)=\alpha_1(\omega_1)+\alpha_2(\omega_2)$. We also plot the variations of the two-photon polarizabilities with $\omega_1$ (in a.u.) for the clock states in Fig \ref{fig4}. These plots reveal the values of the two-photon magic wavelengths which are shown in the figure. Few of these magic wavelengths with high polarizability values are appeared in the optical region and hence can be significant for experiments. A comparison between Table~\ref{3} and Table~\ref{9} reveals availability of more number of magic wavelengths at the optical region in the two-photon case compared to the single-photon case. Therefore, one can have more number of choices in conducting the field free clock experiments at the cost of one more laser source.  

\begin{table} [htb]
\centering
\caption{Two-photon magic wavelengths ($\lambda_{m1}$ and $\lambda_{m2}$) in nm with two-photon polarizability values ($\alpha_{tp}$) in a.u. for the 6$^2S_\frac{1}{2}$ $-$ 5$^2D_{\frac{3}{2},\frac{5}{2}}$ clock transitions of Ba$^+$ ion. The intensities of the two laser beams are considered to be same at the position of Ba$^+$.}
\begin{tabular}{c r r r r r r r r r r r r}
\hline
\multicolumn{6}{c}{6 $^2S_\frac{1}{2}$ $-$ 5 $^2D_\frac{3}{2}$}&& \multicolumn{6}{c}{6 $^2S_\frac{1}{2}$ $-$ 5 $^2D_\frac{5}{2}$}\\
		\cline{1-6} \cline{8-13}
		(${J_v}$,$M_{J_v}$) & $\omega_1$ & $\omega_2$ & $\alpha_{tp}$ & $\lambda_{m1}$ & $\lambda_{m2}$ && (${J_v}$,$M_{J_v}$) & $\omega_1$ &$\omega_2$ & $\alpha_{tp}$ & $\lambda_{m1}$ & $\lambda_{m2}$\\
		
		 	\cline{1-6} \cline{8-13}

($\frac{3}{2},\frac{3}{2}$)	
&	0.055024	&	0.077231	&	528.21	&	828.06	&	589.96	&&($\frac{5}{2},\frac{5}{2}$)&	0.068554	&	0.094410	&	51.02	&	664.64	&	482.61	\\
&	0.072209	&	0.094416	&	85.56	&	630.99	&	482.58	&&&	0.084010	&	0.109867	&	53.61	&	542.35	&	414.72	\\
&	0.085568	&	0.107774	&	12.66	&	532.48	&	422.77	&&&	0.095452	&	0.121308	&	56.25	&	477.34	&	375.60	\\
&	0.095519	&	0.117726	&	36.17	&	477.01	&	387.03	&&&	0.174544	&	0.200400	&	-68.45	&	261.04	&	227.36	\\
&	0.181074	&	0.203281	&	-61.64	&	251.63	&	224.14&&($\frac{5}{2},\frac{3}{2}$)&	0.044445	&	0.070301	&	424.78	&	1025.16	&	648.12	\\
($\frac{3}{2},\frac{1}{2}$)&	0.041102	&	0.063309	&	374.78	&	1108.54	&	719.70	&&&	0.068954	&	0.094810	&	255.99	&	660.78	&	480.57	\\
&	0.070577	&	0.092784	&	-3433.52	&	645.58	&	491.07	&&&	0.082605	&	0.108462	&	-73.92	&	551.58	&	420.09	\\
&	0.085000	&	0.107207	&	-63.55	&	536.04	&	425.00	&&&	0.095360	&	0.121216	&	19.86	&	477.80	&	375.89	\\
&	0.095412	&	0.117619	&	-6.14	&	477.54	&	387.38	&&&	0.176783	&	0.202639	&	-65.40	&	257.74	&	224.85	\\
&	0.182695	&	0.204902	&	-59.68	&	249.40	&	222.37		&&($\frac{5}{2},\frac{1}{2}$)&	0.040841	&	0.066697	&	393.14	&	1115.64	&	683.14	\\
&		&		&		&		&		&&&	0.069232	&	0.095088	&	378.05	&	658.13	&	479.17	\\
&		&		&		&		&		&&&	0.081702	&	0.107559	&	-163.43	&	557.68	&	423.61	\\
&		&		&		&		&		&&&	0.095314	&	0.121170	&	1.45	&	478.03	&	376.03	\\
&		&		&		&		&		&&&	0.177112	&	0.202968	&	-64.97	&	257.26	&	224.49	\\
\hline
\end{tabular}
\label{9}
\end{table}
\clearpage

\subsection{Uncertainty in the estimated magic-wavelengths}
To estimate an approximate uncertainty in the magic wavelength values, we follow two procedures to estimate the maximum deviations of our computed magic wavelengths. Firstly, we consider Table~\ref{3} where our calculated magic wavelengths values are compared with the values of Ref. \cite{Kaur2015} and Ref. \cite{Jiang2021}. It is evident that our magic wavelengths agree excellently (maximum deviation of 0.1\%) with the corresponding wavelengths of Ref. \cite{Kaur2015} and Ref. \cite{Jiang2021} when one of the two intersecting curves, as shown in the Fig.~\ref{fig3}, is almost vertical at the point of intersection. We indicate this type of magic wavelengths as category-I. On the other hand, the magic wavelengths 754.98 nm, 664.23 nm, and 713.69 nm, which appear on the intersections of almost horizontally flat portions of the curves, differ by a maximum amount of 0.8\% from the corresponding magic wavelengths of Ref. \cite{Jiang2021} and 1.7\% from the corresponding magic wavelengths of Ref. \cite{Kaur2015}. We indicate this later type of magic wavelengths as category-II. This comparison shows that category-I magic wavelengths are expected to be more accurate than category-II magic wavelengths. In the second procedure, we reevaluate all the single-photon magic wavelengths by replacing our RCCSD(T) matrix elements with the SD matrix elements \cite{Safronova2010, Barrett2019}. Then we further reevaluate these magic wavelengths by replacing our RCCSD(T) matrix elements with the RCCSD(T) matrix elements of Sahoo {\it et al.} \cite{Saho2009}. These re-evaluated magic-wavelengths using both these approaches differ by a maximum amount of 0.2\% from our presented magic wavelengths in Table~\ref{3}. Therefore, following both the procedures, one can say that the maximum deviation for the category-I magic wavelengths can be 0.2\%, whereas the maximum deviation for the category-II magic wavelengths can be 1.7\%. Consequently, we may conclude that our calculated single-photon category-I magic wavelengths may not have uncertainty by more than 0.5\% and category-II magic wavelengths may not have uncertainty by more than 2\%. These estimates of uncertainty for both these categories can also be valid for the two-photon magic wavelengths in Table~\ref{9}, as similar approaches are followed to evaluate both the single-photon and two-photon magic wavelengths.

\subsection{Application of two-photon polarizability: Spin-mixing in $^{137}$Ba$^+$$-$$^{87}$Rb mixture}
\begin{figure}[htb]
\centering
\subfloat{\includegraphics[width=0.52\linewidth]{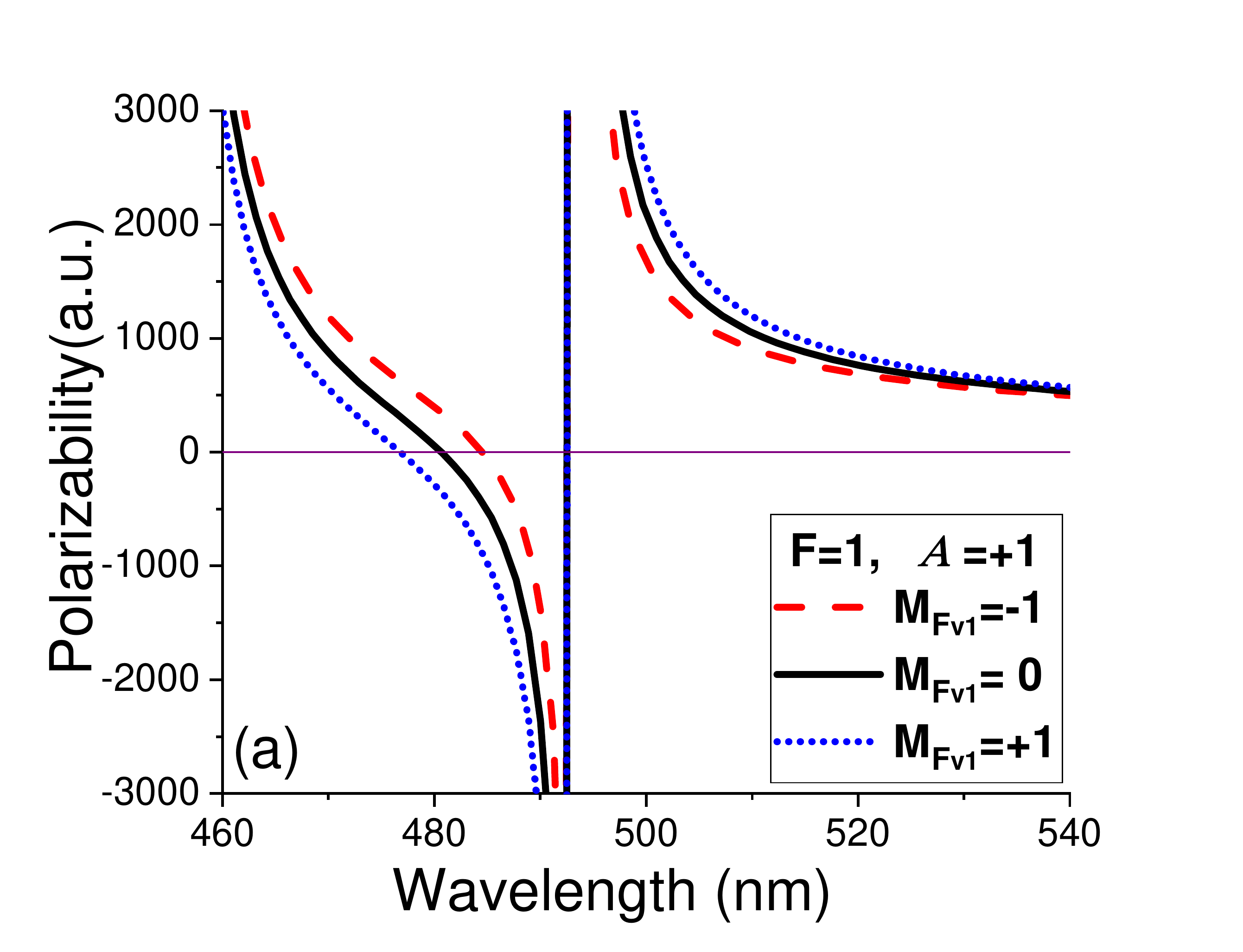}}
\subfloat{\includegraphics[width=0.52\linewidth]{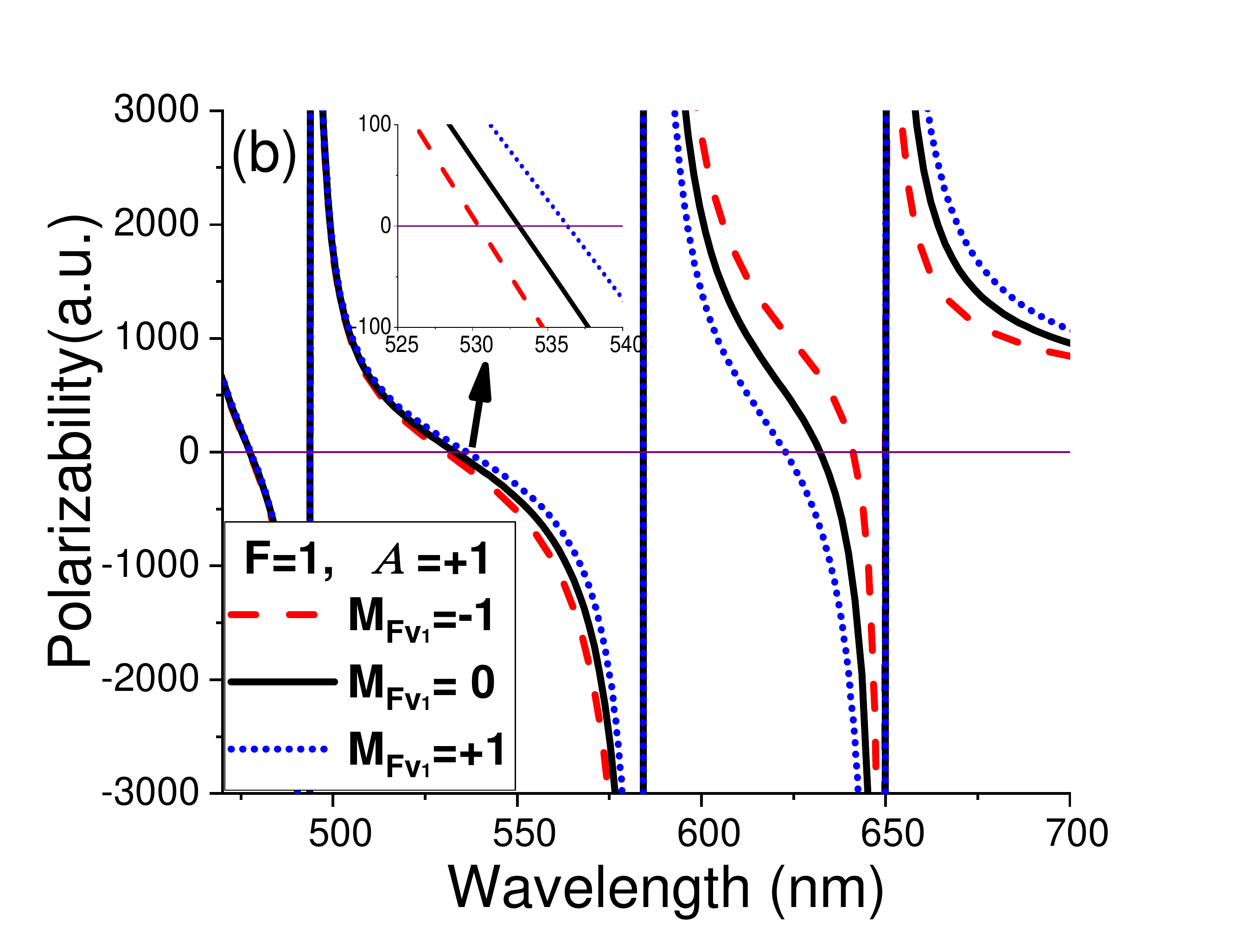}}\\
\caption{ Variations of (a) single-photon and (b) two-photon polarizabilities with wavelength for the $6^2S_{1/2} (F_{v1}=1, M_{F_{v1}}=0, \pm1)$ state of $^{137}$Ba$^+$ ion. In the single-photon case, we assume a left circularly polarized light. In the two-photon case, we assume a left circularly polarized light and a linearly polarized light. In the inset of the panel in (b), we display the enlarged version of the plot around the wavelength 533.08 nm. }
\label{fig5}
\end{figure}

\begin{figure}[htb]
\begin{center}
\includegraphics[width=0.8\textwidth]{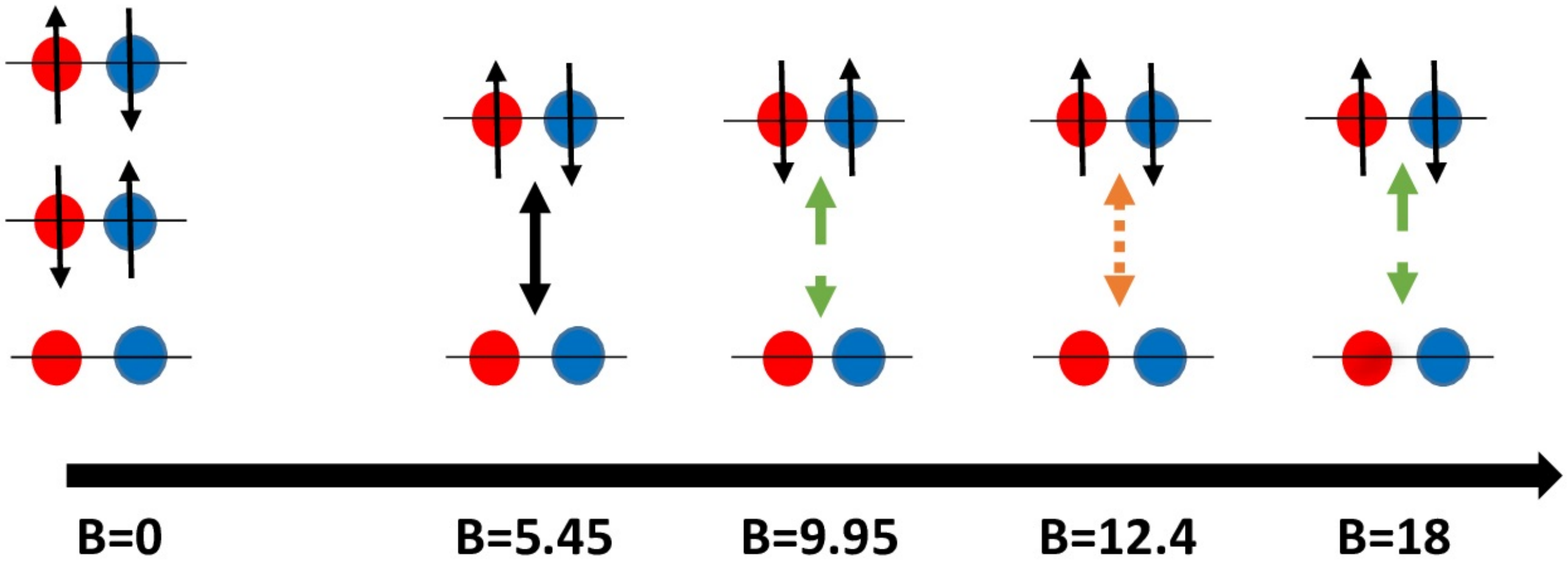}
\caption{Spin-oscillation processes between the spin states $|M_{F_{v1}},M_{F_{v2}}\rangle$ for the $^{137}$Ba$^+$ ion and $^{87}$Rb atom mixture are shown schematically at different magnetic fields (in Gauss). The red ball (on the left) and blue ball (on the right) represent $^{137}$Ba$^+$ ion and $^{87}$Rb atom, respectively. The balls with upward, downward, and no arrows represent $|+1\rangle$, $|-1\rangle$, and $|0\rangle$ spin states, respectively, for both the atom and the ion. Solid black, orange dotted, and green dashed arrows represent spin-changing processes for no-light, single-photon, and two-photon, respectively, cases. In the single-photon case, we assume a left circularly polarized light. In the two-photon case, we assume a left circularly polarized light and a linearly polarized light.}
\label{fig1a}
\end{center}
\end{figure}

 As an exclusive application of the presented theory of dipole polarizability for a two-photon mechanism, we investigate the importance of this polarizability on an ion-atom mixture of a heteronuclear binary system. We consider a mixture of $^{137}$Ba$^+$ and $^{87}$Rb with their hyperfine ground states $6^2S_{\frac{1}{2}}$ $(F_{v1}=1, M_{F_{v1}})$ and $5^2S_{\frac{1}{2}}$ $(F_{v2}=1, M_{F_{v2}})$, respectively. According to the multichannel quantum defect theory calculation, this particular choice of ion-atom pair can be made as a promising candidate for ultracold chemistry \cite{Huber2014, Silva2015, Krych2011}. The composite sublevels (spin states) consisting of the magnetic sublevels of the hyperfine ground states of the ion and the atom are represented by $|M_{F_{v1}}, M_{F_{v2}}\rangle$. The magnetic sublevels of the atom or ion are degenerate in the absence of an external magnetic field or a circularly polarized light. A circularly polarized light can create additional Stark-shifts in opposite directions to $+M_{F_{vn}}$ and $-M_{F_{vn}}$ levels ($n=$ 1 or 2) by inducing the vector part of the polarizability and hence can break this degeneracy completely. In general, this vector part can not be induced by a linearly polarized light. The additional energy shifts to the atom or ion provide a fictitious magnetic field at the tune-out wavelength of light for $M_{F_{vn}}=0$ state \cite{Kien2013}. The tune-out wavelength for a state is the wavelength of the externally applied laser light for which the dipole polarizability of the state becomes zero. However, for a two-photon process, it is an appropriate combination of the wavelengths of the two laser lights for which the two-photon polarizability of the state vanishes. For instance, we can extract the single-photon (480.61 nm) and two-photon (632.21 nm, 533.08 nm) tune-out wavelengths for the 6$^2S_{1/2}$($F_{v1}=1,M_{F_{v1}}=0$) state of $^{137}$Ba$^+$ from Fig. \ref{fig5}. Here it is to be mentioned that the factors to convert the expressions of dynamic polarizabilities at the fine-structure sublevels to the expressions of corresponding polarizabilities at the hyperfine sublevels are represented in the ``Appendix''. Nevertheless, the spin states for the degenerate mixture of $^{137}$Ba$^+$ and $^{87}$Rb are $|0,0\rangle$, $|+1,-1\rangle$, and $|-1,+1\rangle$ (see Fig. \ref{fig1a}) conserving the total $M_F$ value (here we ignore the spin-spin interaction between the ion and the atom). However, if an arbitrary magnetic field is applied to this binary mixture externally, the field may break this degeneracy. Our first target is to investigate whether one can achieve the spin-oscillation or spin-mixing \cite{Widera2005} processes $|0,0\rangle$ $\leftrightarrow$ $|+1,-1\rangle$ and $|0,0\rangle$ $\leftrightarrow$ $|-1,+1\rangle$ by maintaining the required degeneracy at some non-zero magnetic fields. After that, we investigate the possibilities of these spin-oscillation processes in the presence of a single laser beam and then two counter-propagating laser beams having appropriate polarizations.

\begin{figure}[htb]
\centering
{\includegraphics[width=0.50\linewidth]{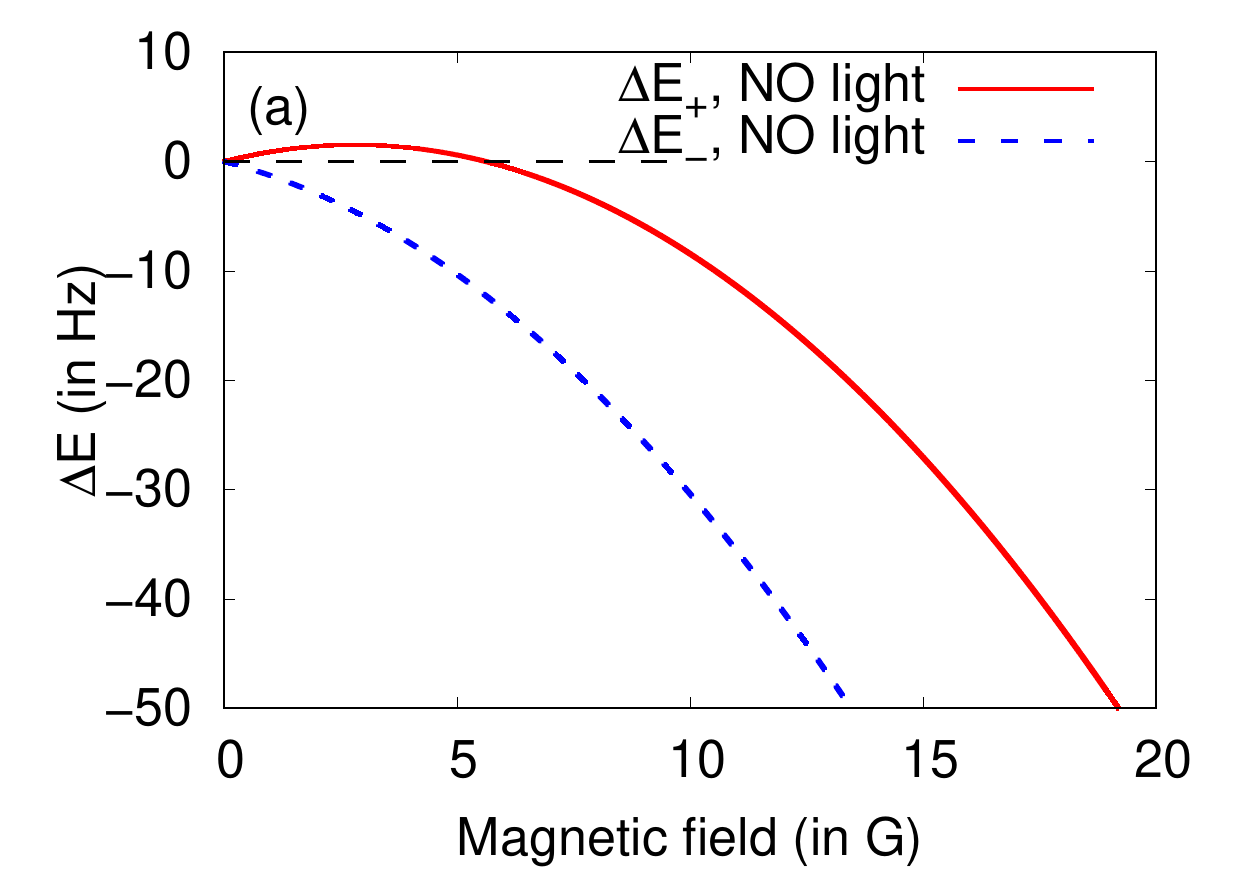}}\\
{\includegraphics[width=0.50\linewidth]{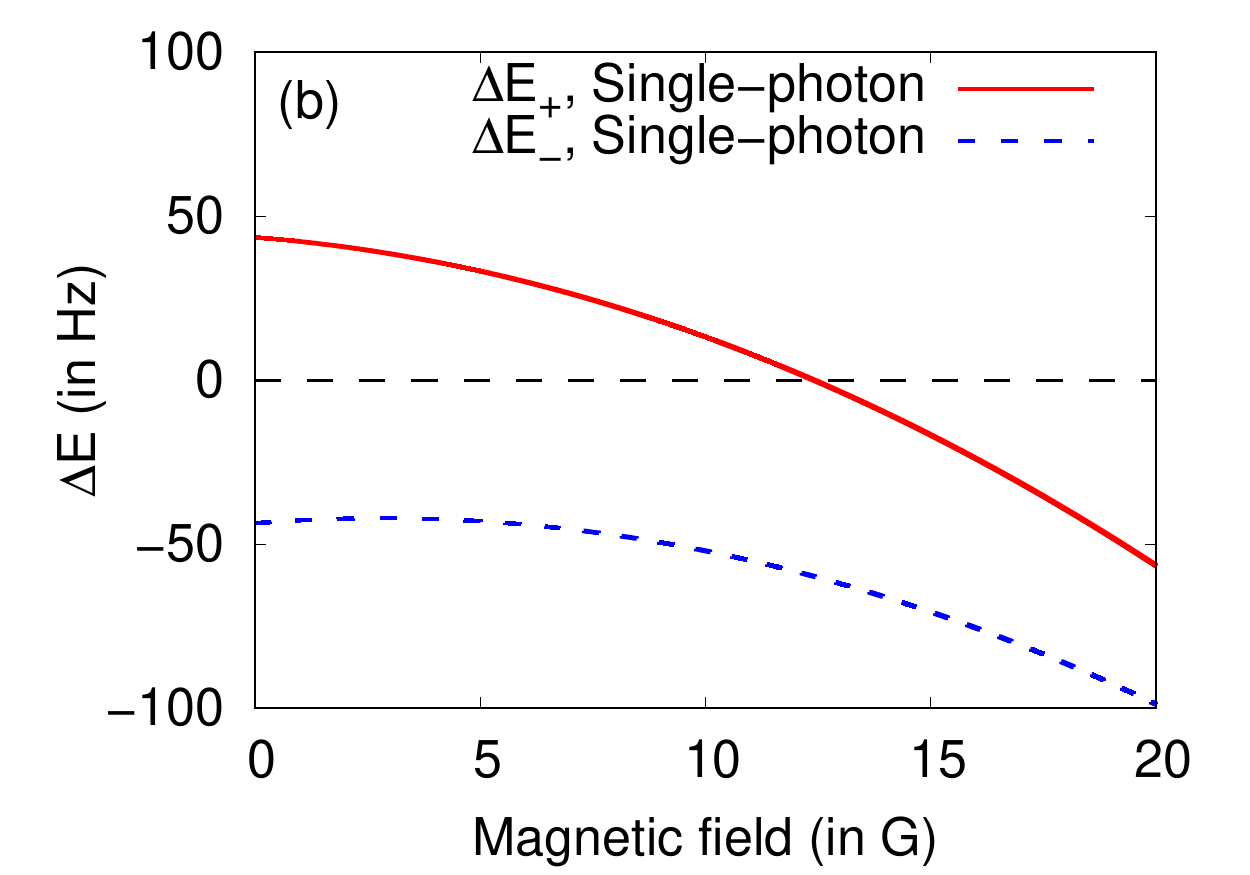}}\\
{\includegraphics[width=0.50\linewidth]{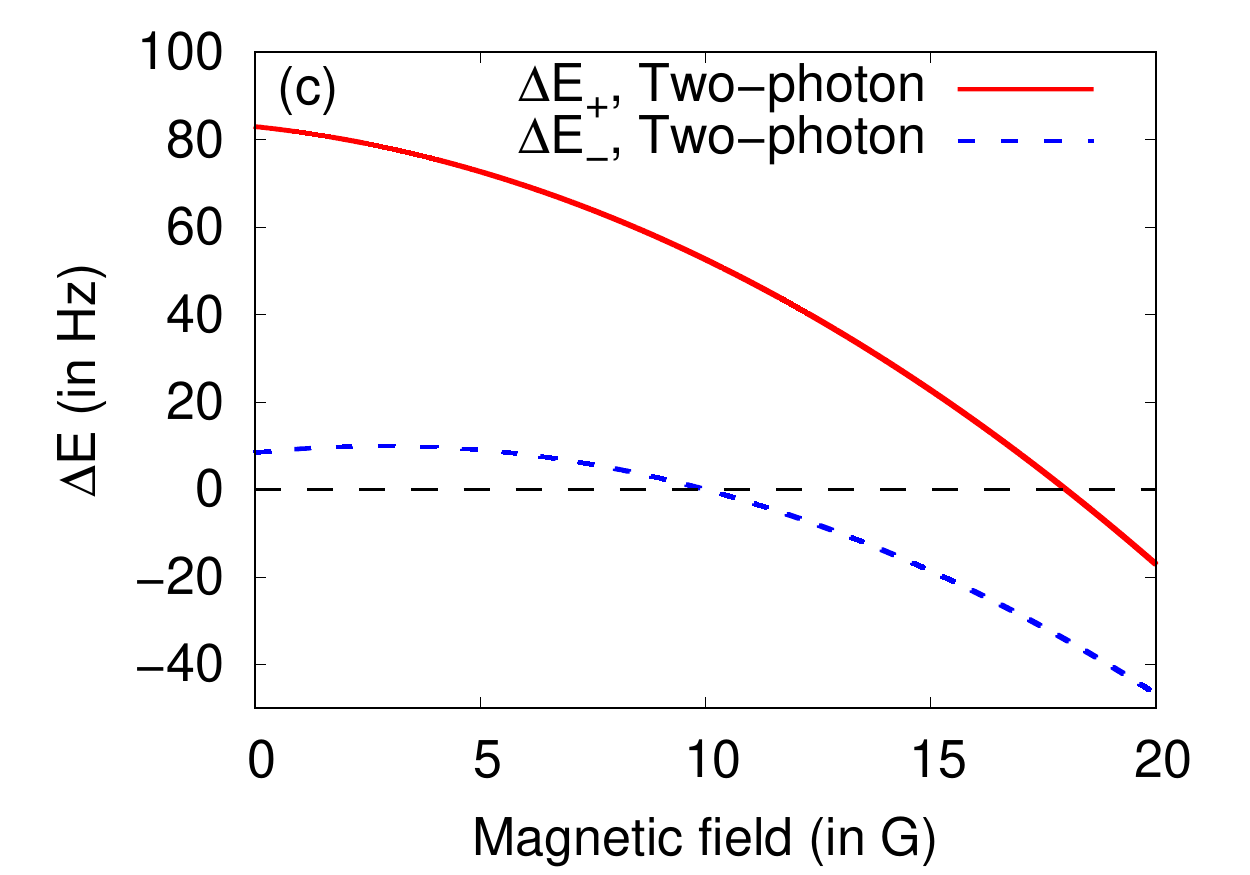}}\\
\caption{Magnetic energy diagram for the two heteronuclear spin-oscillation processes in the spin-1 $^{137}$Ba$^+$ and $^{87}$Rb mixture (a) without light shift, (b) with light shift by single-photon interaction ($\lambda =480.61$ nm), and (c) with light shift by two-photon interaction (632.21 nm, 493.30 nm) and (533.08 nm, 423.14 nm).}
\label{fig6}
\end{figure}

We indicate the energy difference between the composite levels by $\Delta E_+(B)$ =$ E^{|0,0\rangle}- E^{|+1,-1\rangle}$ and $\Delta E_{-}(B)$ = $E^{|0,0\rangle} -E^{|-1,+1\rangle}$. The magnetic energy $E^{|M_{F_{v1}}, M_{F_{v2}}\rangle}(B)$ for the mixture of  $^{137}$Ba$^+$ and $^{87}$Rb is determined using the Hamiltonian $\hat{H}_Z=-\zeta_1 \hat{B}_Z+\zeta_2\hat{B}_Z^2$, where $\zeta_1$ and $\zeta_2$ are the linear and quadratic Zeeman shifts, respectively. $\hat{B}_Z$ represents the $z$-component of the external magnetic field. The coefficients  $\zeta_1$ and $\zeta_2$ can be evaluated from the power series expansion of the  Breit-Rabi formula \cite{Vanier1988, Zhang2005}. In this analysis, we consider the ground state magnetic dipole hyperfine coupling constant for $^{137}$Ba$^+$ as 4.0189 GHz \cite{Trapp2000}. Nevertheless, in the absence of any light, as we increase the magnetic field from zero, the values of $\Delta E_{\pm}(B)$ shift from zero as shown in Fig. \ref{fig6}(a). $\Delta E_+(B)$ becomes zero again at $B$ =5.45 G indicating the fact that spin-oscillation is also possible between $|0,0\rangle$ and $|+1,-1\rangle$ states at some non-zero magnetic field (see also Fig. \ref{fig1a}). However, the other spin-oscillation, $|0,0\rangle \leftrightarrow |-1,+1\rangle$, does not seem to be taken place at a finite non-zero magnetic field.

Fig. \ref{fig6}(b) indicates that the two spin-mixing processes cannot occur at zero magnetic field when the binary system is exposed under a left circularly polarized light  with wavelength 480.61 nm. Here we consider the intensity of the external light beam is 10 W/m$^2$. The tune-out wavelength 480.61 nm is very close to the first resonance line of $^{137}$Ba$^+$ ion and far away from the resonance lines of $^{87}$Rb atom. Therefore, the light affects the $^{137}$Ba$^+$ ion but keeps the $^{87}$Rb atom almost unaffected. As the light is circularly polarized, the dipole polarizability induced by this light has a non-zero vector part. This vector part leads to a fictitious magnetic field felt only by the $^{137}$Ba$^+$ ion. Fig. \ref{fig6}(b) reveals that  $\Delta E_{-}(B)$ and $\Delta E_{+}(B)$ split away for this fictitious magnetic field when the external magnetic field is zero. Therefore, the $|0,0\rangle \leftrightarrow |-1,+1\rangle$ spin-oscillation is suppressed forever as the external magnetic field increases from 0. Whereas, the spin oscillation $|0,0\rangle \leftrightarrow |+1,-1\rangle$ becomes feasible at the magnetic field of 12.40 G (see also Fig. \ref{fig1a}). This shift of external magnetic field from 5.45 G (without light) to 12.40 G (single-photon interaction) is a manifestation of the vector polarizability created by the circularly polarized light. Therefore, by tuning the light to an appropriate wavelength, one can switch-on the spin-oscillation $|0,0\rangle \leftrightarrow |+1,-1\rangle$ at a desired non-zero magnetic field. Nevertheless, the spin-oscillation $|0,0\rangle \leftrightarrow |-1,+1\rangle$ can not be achieved unless we change the direction of circular polarization of the light.

In the two-photon case, one of the laser lights with with frequency $\omega_1$ (wavelength $\lambda_1$) is considered to have left circular polarization and another with frequency $\omega_2$ (wavelength $\lambda_2$) is considered to have linear polarization.. Both the laser lights have intensities of 10 W/m$^2$. The frequency difference between the two lights is considered to be $\Delta\omega=0.022206903$ a.u. which corresponds to the frequency of $6^2S_{\frac{1}{2}}$ -- $5^2D_{\frac{3}{2}}$ transition. Here we get two sets of tune-out wavelengths at ($\lambda_{1},\lambda_{2}$) = (632.21 nm, 493.30 nm) and (533.08 nm, 423.14 nm) for the 6$^2S_\frac{1}{2}(M_{F_{v1}}=0)$ state of $^{137}$Ba$^+$ as depicted in Fig. \ref{fig5}(b). Here also, $^{87}$Rb atom behaves as almost transparent for these two sets of tune-out wavelengths. Fig. \ref{fig6}(c) shows the external magnetic field variation of $\Delta E_{+}(B)$=$E^{|0,0\rangle}-E^{|+1,-1\rangle}+\delta E_{l}^{TP}$ for lights with wavelengths (632.21 nm, 493.30 nm) and $\Delta E_{-}(B)$=$E^{|0,0\rangle}-E^{|-1,+1\rangle}+\delta E_{l}^{TP}$ for lights with wavelengths (533.08 nm, 423.14 nm). Here $\delta E_{l}^{TP}$ indicates the light shift due to the two-photon interaction. This figure shows that by proper choice of wavelengths of the lights and the external magnetic field (9.95 G), $\Delta E_{-}{(B)}$ can be made zero to switch on the $|0,0\rangle \leftrightarrow |-1,+1\rangle$ spin-oscillation. This particular spin-oscillation is not observed in no light condition and the single-photon case with left circularly polarized light. In order to understand its possibility in a two-photon case, one needs to look Fig. \ref{fig5}(b). This figure shows that the two-photon polarizabilities of $M_{F_{v1}}=+1$ and $M_{F_{v1}}=-1$ for $^{137}$Ba$^+$ ion are positive and negative, respectively, at $\lambda_{1}=533.08$ nm. However, the signs of these polarizabilities are reversed at the other wavelength $\lambda_{1}=632.21$ nm. Therefore, the signs of the fictitious magnetic fields are changed accordingly between these two wavelengths. As a consequence, the spin-oscillation $|0,0\rangle \leftrightarrow |-1,+1\rangle$ is achievable at 9.95 G for $\lambda_{1}=533.08$ nm and $\lambda_{2}=423.14$ nm, whereas the spin-oscillation $|0,0\rangle \leftrightarrow |+1,-1\rangle$ can occur at 18 G for $\lambda_{1}=632.21$ nm and $\lambda_{2}=493.30$ nm (see also Fig. \ref{fig1a}). On this account, we can say that by changing the wavelengths of the two counter-propagating lights, one can control both the spin-oscillation processes of the heteronuclear spin-1 mixture of  $^{137}$Ba$^+$ and $^{87}$Rb. It is to be mentioned here that both these spin-oscillation processes can also be achieved at different magnetic fields if we consider $\Delta\omega=0.025856323$ a.u. which corresponds to the frequency of $6^2S_{\frac{1}{2}}$ -- $5^2D_{\frac{5}{2}}$ transition.
    
\section{Conclusion}\label{CON}
This work shows the frequency or wavelength dependence of dipole polarizabilities of a few low-lying states of singly ionized barium atom. These low-lying states, $6^2S_{\frac{1}{2}}$, $5^2D_{\frac{3}{2}}$, and $5^2D_{\frac{5}{2}}$, are associated with the proposed optical clock transitions $6^2S_{\frac{1}{2}}$ -- $5^2D_{\frac{3}{2},\frac{5}{2}}$ of Ba$^{+}$. From the profiles of these polarizabilities, we extract the magic wavelengths for the zero differential Stark-shifts in the clock transitions due to the single-photon and two-photon interactions. The single-photon differential polarizabilities and the magic wavelength values are compared with the relevant other theoretical and experimental values to calibrate our calculations. The excellent agreements in this comparison can give an indication of the accuracy of our two-photon calculations which are the primary objective of this work. We find a few two-photon magic wavelengths in the optical region providing high polarizability values to the clock states. These magic wavelengths can be advantageous in conducting clock experiments on Ba$^{+}$ with minimal errors. 

Besides the clock-related experiments, the most important advantage of using a two-photon mechanism is highlighted here in the spin-oscillation processes of a spin-1 mixture of $^{137}$Ba$^{+}$ ion and $^{87}$Rb atom in the presence of an external magnetic field. The spin-states of this binary mixture, $|0,0\rangle$, $|+1,-1\rangle$ and $|-1,+1\rangle$, act as degenerate states in no light and no magnetic field conditions. However, when a single circularly polarized laser light is applied, this degeneracy is lifted completely. In this situation, a non-zero magnetic field is required to maintain the degeneracy between the $|0,0\rangle$ state and one of the two states, $|+1,-1\rangle$ and $|-1,+1\rangle$, depending on the direction of circular polarization. Consequently, such a magnetic field can activate only one of the two spin-oscillation processes, $|0,0\rangle$ $\leftrightarrow$ $|+1,-1\rangle$ and $|0,0\rangle$ $\leftrightarrow$ $|-1,+1\rangle$. In this work, we have demonstrated that a two-photon interaction using a circularly polarized light and a linearly polarized light can activate both the spin-oscillation processes without changing the direction of the circular polarization. However, each of these processes needs an appropriate combination of wavelengths of the lights and accordingly, an appropriate external magnetic field. The triggering of both the spin oscillation processes by tuning the wavelength of the two laser lights and the strength of the magnetic field is a manifestation of the fictitious magnetic field generated by the vector component of the dipole polarizability. .  

Therefore, the two-photon model can induce coherent heteronuclear spin-mixing dynamics and significantly influence the quantum phases of the binary-species spin-1 BEC. Our model can be applied to other atomic mixtures to generate entanglement between two different atoms  \cite{Luo2017, Zou2018} and control the singlet-pairing process \cite{Jie2021}.

\appendix*
\section{Dynamic valence polarizability at the hyperfine level}
The scalar $\alpha_{v_F}^{(s)}(\omega)$ , vector $\alpha_{v_F}^{(1)}(\omega)$ and tensor $\alpha_{v_F}^{(2)}(\omega)$ components of valence polarizability at a hyperfine level can be related to the corresponding polarizability at the fine-structure level by \cite{Beloy2009,Dzuba2010,Das2020}
\begin{equation}
\alpha_{v_F}^{(s)}(\omega)=\alpha_{v}^{(s)}(\omega)
\end{equation}
\begin{equation}
\alpha_{v_F}^{(1)}(\omega)=(-1)^{J_v+F_v+I+1}\left\{\begin{array}{ccc} F_v & J_v & I\\ J_v & F_v& 1 \end{array}\right \}\sqrt{\frac{F_v(2F_v+1)(2J_v+1)(J_v+1)}{J_v(F_v+1)}}\alpha_v^{(1)}(\omega)
\end{equation}
and
\begin{eqnarray}
\alpha_{v_F}^{(2)}(\omega)=(-1)^{J_v+F_v+I}\left\{\begin{array}{ccc} F_v & J_v & I\\ J_v & F_v& 2 \end{array}\right \}&& \sqrt{\left(\frac{F_v(2F_v-1)(2F_v+1)}{(2F_v+3)(F_v+1)}\right)} \times \nonumber\\ 
&&\sqrt{ \left(\frac{(2J_v+3)(2J_v+1)(J_v+1)}{J_v(2J_v-1)}\right)}\alpha_v^{(2)}(\omega).
\end{eqnarray}
Here $F_v$ (\textit{$\textbf{F}_v =\textbf{J}_v+\textbf{I} $}) and $M_{F_v}$ represent hyperfine quantum number and its magnetic component, respectively. Hence the valence polarizability $\alpha^V_{v_F}(\omega)$ at a hyperfine level can be written as
\begin{eqnarray}\label{4}
\alpha^V_{v_F}(\omega)= \alpha_{v_F}^{(s)}(\omega) + A\frac{M_{F_v}}{2F_v} \alpha_{v_F}^{(1)}(\omega) + C_1\alpha_{v_F}^{(2)}(\omega),
\end{eqnarray}
where 
\begin{equation}
C_1=\frac{3M_{F_v}^2-J_v(F_v+1)}{F_v(2F_v-1)}\ \ \text{for}\ \ \ 
F_v>\frac{1}{2}.
\end{equation}


\begin{thebibliography}{}

\bibitem{Arnold2020}	
K. J. Arnold, R. Kaewuam, S. R. Chanu, T. R. Tan, Z. Zhang , and M. D. Barrett, Phys. Rev. Lett. \textbf{124}, 193001 (2020).
\bibitem{Chanu2020}	
S. Chanu, V. Koh, K. Arnold, R. Kaewuam, T. Tan, Z. Zhang, M. Safronova, and M. Barrett, Phys. Rev. A \textbf{101}, 042507 (2020).
\bibitem{Inlek2017}
I. V. Inlek, C. Crocker, M. Lichtman, K. Sosnova, and C. Monroe, Phys. Rev. Lett. \textbf{118}, 250502 (2017).
\bibitem{Hucul2017}
D. Hucul, J. E. Christensen, E. R. Hudson, and W. C. Campbell, Phys. Rev. Lett. \textbf{119}, 100501 (2017).
\bibitem{Dutta2014}
N. N. Dutta and S. Majumder, Phys. Rev. A \textbf{90}, 012522 (2014).
\bibitem{Kozlov2018}
M. G. Kozlov, M. S. Safronova, J. R. Crespo López-Urrutia, and P. O. Schmidt, Rev. Mod. Phys. \textbf{90}, 045005 (2018).
\bibitem{Munshi2015}
D. De Munshi, T. Dutta, R. Rebhi, and M. Mukherjee, Phys. Rev. A\textbf{ 91}, 040501(R) (2015).
\bibitem{Dijck2015}
E. A. Dijck, M. Nunez Portela, A. T. Grier,  K. Jungmann, A. Mohanty, N. Valappol, and L. Willmann, Phys. Rev. A \textbf{91}, 060501(R) (2015).
\bibitem{Zhang2020}
Z. Zhang, K. J. Arnold, S. R. Chanu, R. Kaewuam, M. S. Safronova, and M. D. Barrett, Phys. Rev. A \textbf{101}, 062515 (2020).
\bibitem{Tang2013}
Y.-B. Tang, H.-X. Qiao, T.-Y. Shi, and J. Mitroy, Phys. Rev. A {\textbf 87}, 042517 (2013).
\bibitem{Mitroy2010}
J. Mitroy, M. S. Safronova, and C. W. Clark, J. Phys. B \textbf{43}, 202001 (2010).
\bibitem{Gallagher1979}
T. F. Gallagher and W. E. Cooke, Phys. Rev. Lett. {\textbf 42}, 835 (1979).
\bibitem{Porsev2006}
S. G. Porsev and A. Derevianko, Phys. Rev. A  {\textbf 74}, 020502(R) (2006).
\bibitem{Jefferts2014}
S. R. Jefferts, T. P. Heavner, T. E. Parker. J. H. Shirley, E. A. Donely, N. Ashby, F. Levi, D. Calonico, and G. A. Costanzo, Phys. Rev. Lett. \textbf{112}, 050801 (2014).
\bibitem{Gerginov2018}
V. Gerginov and K. Beloy, Phys. Rev. Appl. \textbf{10}, 014031 (2018).
\bibitem{Martin2019}
K. W. Martin, B. Stuhl, J. Eugenio, M. S. Safronova, G. Phelps, J. H. Burke, and N. D. Lemke, Phys. Rev. A \textbf{100}, 023417 (2019).
\bibitem{Perrella2019}
C. Perrella, P.S. Light, J.D. Anstie, F.N. Baynes, R.T. White, and A.N. Luiten,
Phys. Rev. Appl. \textbf{12}, 054063 (2019).
\bibitem{Jackson2019}
S. Jackson and A. C. Vutha, Phys. Rev. A \textbf{99}, 063422 (2019).
\bibitem{Martin2018}
K. W. Martin, G. Phelps, N. D. Lemke, M. S. Bigelow, B. Stuhl, M. Wojcik, M. Holt, I. Coddington, M. W. Bishop, and J. H. Burke,
Phys. Rev. Applied \textbf{9}, 014019 (2018).
\bibitem{Hall1989}
J. L. Hall, M. Zhu, and P. Buch, JOSA B \textbf{6}, 2194 (1989).
\bibitem{Marian2004}
A. Marian, M. C. Stowe, J. R. Lawall, D. Felinto, J. Ye, Science \textbf{306}, 2063 (2004).
\bibitem{diddams2020}
S. A. Diddams, K. Vahala, T. Udem,
Science \textbf{369}, 6501 (2020).
\bibitem{Kang2016}
YH. Kang, Y. Xia, and PM. Lu, Quantum Inf Process \textbf{15}, 4521–4535 (2016).
\bibitem{Dutta2015}
N. N. Dutta, S. Roy, and P. C. Deshmukh, Phys. Rev. A \textbf{92},
052510 (2015).
\bibitem{Xu2012}
Z. F. Xu, D. J. Wang, and L. You, 
Phys. Rev. A  \textbf{86},  013632 (2012).
\bibitem{Zhang2011}
J. Zhang, T. Li, and Y. Zhang, 
Phys. Rev. A \textbf{83},  023614 (2011).
\bibitem{Li2017}
Z. B. Li, Y. M. Liu, D. X. Yao, and C. G. Bao, 
Journal of Physics B: Atomic, Molecular and Optical
Physics \textbf{50}, 135301 (2017).
\bibitem{Xu2010}
Z. F. Xu, J. Zhang, Y. Zhang, and L. You,  Phys. Rev. A \textbf{81}, 033603 (2010).
\bibitem{Modgno2002}
G. Modugno, M. Modugno, F. Riboli, G. Roati, and M. Inguscio, Phys.  Rev.  Lett. \textbf{89}, 190404 (2002).
\bibitem{Thalhammer2008}
G. Thalhammer, G. Barontini, L. De Sarlo, J. Catani, F. Minardi, M. Inguscio, Phy. Rev. Lett. \textbf{100}, 210402 (2008).
\bibitem{McCarron2011}
D. J. McCarron, H. W. Cho, D. L. Jenkin, M. P. K\"oppinger, and S. L. Cornish, Phys. Rev. A \textbf{84}, 011603 (2011).
\bibitem{ARoy2015}
A. Roy and D. Angom, Phy. Rev. A \textbf{92}, 011601(R) (2015).
\bibitem{Li2015}
X. Li, B. Zhu, X. He, F. Wang, M. Guo, Z. Xu,
S. Zhang, and D. Wang, Phys. Rev.  Lett. \textbf{114}, 255301 (2015).
\bibitem{Mil2020}
A. Mil, T . V. Zache, A. Hegde, A. Xia, R. P. Bhatt, M. K. Oberthaler,  P. Hauke, J. Berges,  and F.  Jendrzejewski, Science \textbf{367}, 1128 (2020).
\bibitem{Chen2018}
J-J Chen, Z-F Xu, and L You,
Phys. Rev. A, \textbf{98}, 023601 (2018).
\bibitem{Li2020}
L. Li, B. Zhu, B. Lu, S. Zhang, and D. Wang,
Phys. Rev. A,  \textbf{101}, 053611 (2020). 
\bibitem{Fang2020}
F. Fang, S. Wu, A. Smull, J. A. Isaacs, Y. Wang, C. H. Greene, and D. M. Stamper-Kurn,  Phys. Rev. A \textbf{101}, 012703 (2020).



\bibitem{Hirzler2020}
H. Hirzler, T. Feldker, H. Fürst, N. V. Ewald, E. Trimby, R. S. Lous, J. D. Arias Espinoza, M. Mazzanti, J. Joger, and R. Gerritsma, Phys. Rev. A \textbf{102}, 033109 2020.
\bibitem{Tomza2019}
M. Tomza, K. Jachymski, R. Gerritsma, A. Negretti, T. Calarco, Z. Idziaszek, and P. S. Julienne,
Rev. Mod. Phys. \textbf{91}, 035001(2019) 


\bibitem{Huber2014}
T.  Huber,  Optical Trapping of Barium Ions-
Towards Ultracold Interactions in Ion-Atom Ensembles,  Thesis (2014).
\bibitem{Krych2011}
M. Krych, W. Skomorowski, F. Pawlowski, R. Moszynski, and Z. Idziaszek, Phys. Rev. A \textbf{83}, 032723 (2011).
\bibitem{Widera2005}
A. Widera, F. Gerbier, S. Fölling, T. Gericke, O. Mandel, and I. Bloch,
Phys. Rev. Lett. \textbf{95}, 190405 (2005).
\bibitem{Bhowmik2020}
A. Bhowmik, N. N. Dutta,  and S. Majumder, Phys. Rev. A \textbf{102}, 063116 (2020).
\bibitem{Das2020}
A. Das, A. Bhowmik, N. N. Dutta, and S. Majumder, Phy. Rev. A \textbf{102}, 012801 (2020).


\bibitem{Chaudhuri2003}
R. K. Chaudhuri, B. K. Sahoo, B. P. Das, H. Merlitz, U. S. Mahapatra, and D. Mukherjee, J. Chem. Phys. {\bf 119}, 10633 (2003).
\bibitem{Dutta2016}
N. N. Dutta and S. Majumder, Indian J. Phys. \textbf{90}, 373 (2016).
\bibitem{Dutta2013}
N. N. Dutta, S. Roy, G. Dixit, and S. Majumder, Phys. Rev. A {\bf 87}, 012501 (2013).
\bibitem{Bartlett2007}
R. J. Bartlett and M. Musial,  Rev. Mod. Phys. \textbf{79}, 291 (2007).
\bibitem{Biswas2018}
S. Biswas, A. Das, A. Bhowmik, and S. Majumder, Mon. Not. R. Astron. Soc. \textbf{477}, 5605 (2018).
\bibitem{Majumder2001}
S. Majumder, G Gopakumar, H. Merlitz, and B. P. Das,
J. Phys. B: At. Mol. Opt. Phys. \textbf{34}, 4821 (2001)
\bibitem{Bhowmik2017}
A. Bhowmik, S. Roy, N. N. Dutta, and S. Majumder, J. Phys. B: At. Mol. Opt. Phys. \textbf{50}, 125005 (2017).
\bibitem{Das2018}
A. Das, A. Bhowmik, N. N. Dutta, and S. Majumder, J. Phys. B: At. Mol. Opt. Phys. \textbf{51}, 025001 (2018).
\bibitem{Lindgren1986}
I. Lindgren, J. Morrison, Atomic Many-Body Theory vol. 3, Springer-Verlag, Berlin, Heidelberg, 1986, https://doi.org/10.1007/978-3-642-61640-2.
\bibitem{Lindgren1985}
I. Lindgren and J. Morrison, Atomic Many-body Theory, ed. G. E. Lambropoulos and H.
Walther(3rd ed.; Berlin: Springer), 3 (1985).
\bibitem{Boyle1998}
J. Boyle, M. Pindzola, Many-Body Atomic Physics, Cambridge University Press, Cambridge, 1998. doi:10.1017/CBO9780511470790.
\bibitem{Johnson1996}
W. R. Johnson, Z. W. Liu, and J. Sapirstein, At. Data Nucl. Data Tables \textbf{64}, 279 (1996).
\bibitem{Bhowmik2018}
A. Bhowmik, N. N. Dutta, and S. Majumder, Phys. Rev. A \textbf{97}, 022511 (2018).
\bibitem{Clementi1990}	
E. Clementi (Ed.), \textit{Modern Techniques in Computational Chemistry: MOTECC-90}, (Springer, Netherlands, 1990).
\bibitem{Roy2015}
S. Roy and S. Majumder, Phy. Rev. A \textbf{92}, 012508 (2015).
\bibitem{Huzinaga1993}
S. Huzinaga and M. Klobukowski, Chem. Phys. Lett. \textbf{212},
260 (1993).
\bibitem{Huzinaga1985}
S. Huzinaga, M. Klobukowski, and H. Tatewaki, Can. J. Chem. {\bf 63}, 1812 (1985).
\bibitem{Safronova2010}
U. I. Safronova, Phys. Rev. A \textbf{81}, 052506 (2010).
\bibitem{Barrett2019}
M. D. Barrett, K. J. Arnold, and M. S. Safronova, Phys. Rev. A \textbf{100}, 043418 (2019).
\bibitem{Sahoo2009}	
B. K. Sahoo, L. W. Wansbeek, K. Jungmann, and R. G. E. Timmermans, Phys. Rev. A \textbf{79}, 052512 (2009).
\bibitem{Jiang2021}
J. Jiang, Y. Ma, X. Wang, C-Z Dong, and Z. W. Wu, Phys. Rev. A \textbf{103}, 032803 (2021).
\bibitem{Dutta2020}
N. N. Dutta, Chem. Phys. Lett. {\bf 758}, 137911 (2020).
\bibitem{NIST2020}
A. Kramida, Yu. Ralchenko, J. Reader, and NIST ASD Team. NIST Atomic Spectra Database (ver. 5.8), [Online]. Available: https://physics.nist.gov/asd. National Institute of Standards and Technology, Gaithersburg, MD. DOI: https://doi.org/10.18434/T4W30F.
\bibitem{Saho2009}
B. K. Sahoo, R. G. E. Timmermans, B. P. Das, and D. Mukherjee, Phys. Rev. A \textbf{80}, 062506 (2009).
\bibitem{Kaur2015}
J. Kaur, S. Singh, B. Arora, and B. K. Sahoo, Phys. Rev. A \textbf{92}, 031402(R) (2015).
\bibitem{Kastberg1993}
A. Kastberg, P. Villemoes, A. Arnesen, F. Heijkenskjöld, A.
Langereis, P. Jungner, and S. Linnaeus, J. Opt. Soc. Am. B \textbf{10},
1330 (1993).
\bibitem{Klose2002}
J. Klose, J. Fuhr, and W. Wiese, J. Phys. Chem. Ref. Data \textbf{31},
217 (2002).
\bibitem{Davidson1992}
M. D. Davidson, L. C. Snoek, H. Volten, and A. Doenszelmann, Astron. Astrophys. \textbf{255}, 457 (1992).
\bibitem{Woods2010}
S. L. Woods, M. E. Hanni, S. R. Lundeen, and E. L. Snow, Phys. Rev. A \textbf{82}, 012506 (2010).
\bibitem{Kurz2008}
N. Kurz, M. R. Dietrich, G. Shu, R. Bowler, J. Salacka, V.
Mirgon, and B. B. Blinov, Phys. Rev. A \textbf{77}, 060501(R) (2008).

\bibitem{Silva2015}
H. Silva, M. Raoult, M. Aymar and O. Dulieu, New J. Phys. \textbf{17}, 045015 (2015).
\bibitem{Kien2013}
F. L. Kien, P. Schneeweiss, and A. Rauschenbeute, Eur. Phys. J.
D \textbf{67}, 92 (2013).


\bibitem{Vanier1988}
J. Vanier and C. Audoin, The Quantum Physics of Atomic Frequency Standards (Hilger, Philadelphia, (1988).
\bibitem{Zhang2005}
W. X. Zhang, D. L. Zhou, M. S. Chang, M. S. Chapman, and L. You, Phys. Rev. A \textbf{72}, 013602 (2005).

\bibitem{Trapp2000}
S. Trapp, G. Marx, G. Tommaseo, A. Klaas, A. Drakoudis, G. Revalde, G. Szawiola, and G. Werth,  
Hyperfine Interactions  \textbf{127}, 57 (2000).
\bibitem{Luo2017} 
X. Y. Luo, Y. Q. Zou, L. N. Wu, Q. Liu, M. F. Han, M. K.
Tey, and L. You, Science \textbf{355},  620(2017).
\bibitem{Zou2018}
Y. Q. Zou, L. N. Wu, Q. Liu, X. Y. Luo, S. F. Guo,
J. H. Cao, M. K. Tey, and L. You,  Proceedings of the National Academy of
Sciences \textbf{115}, 6381 (2018).
\bibitem{Jie2021}
J. Jie, Y. Yu, D. Wang, and P. Zhang, arXiv:2008.10071
\bibitem{Beloy2009}
K. Beloy, Theory of the ac stark effect on the atomic hyperfine structure and applications to microwave atomic clocks, Doctoral dissertation, University of Nevada, Reno, USA, 2009.
\bibitem{Dzuba2010}
V. A. Dzuba, V. V. Flambaum, K. Beloy, and A. Derevianko, Phys. Rev. A \textbf{82}, 062513 (2010).


\end{thebibliography}
\end{document}